\documentclass[11pt,twoside]{article}
\input{epsf.sty}
\usepackage{mathrsfs}
\usepackage{amsmath}
\usepackage{amssymb}
\usepackage{graphicx}
\usepackage{subfigure}
\usepackage{appendix}
\usepackage{cite}
\newtheorem{theorem}{Theorem}
\newtheorem{definition}{Definition}
\newtheorem{proposition}{Proposition}
\newtheorem{lemma}{Lemma}

\numberwithin{equation}{section}
\newcommand*{\QEDB}{\hfill\ensuremath{\square}}

\title{The Ablowitz-Ladik system on a graph}
\author{Baoqiang Xia
\\
School of Mathematics and Statistics, Jiangsu Normal
University,\\
 Xuzhou, Jiangsu 221116, P. R. China,
 \\
 E-mail address:
xiabaoqiang@126.com
}

\date{}
\begin{document}
\maketitle

\begin{abstract}

This paper presents an approach to study initial-boundary value (IBV) problems for integrable nonlinear differential-difference equations (DDEs) posed on a graph.
As an illustrative example, we consider the Ablowitz-Ladik system posed on a graph that is constituted by $N$ semi-infinite lattices (edges)
connected through some boundary conditions.
We first show analyzing this problem is equivalent to analyzing a certain matrix IBV problem;
then we employ the unified transform method (UTM) to analyze this matrix IBV problem.
We also compare our results with some previously known studies.
In particular, we show that the inverse scattering method (ISM) for the integrable DDEs on the integers
can be recovered from the UTM applied to our $N=2$ graph problem as a particular case,
and the nonlocal reductions of integrable DDEs can be obtained as local reductions from our results.

\noindent {\bf Keywords:}\quad Ablowitz-Ladik system, inverse scattering method, unified transform method, initial-boundary value problem.

\end{abstract}
\newpage

\section{ Introduction}

In recent years, the subject of nonlinear evolution equations on graphs has attracted increasing attentions
due to its rich mathematical structures as well as wide physical applications;
see for example \cite{M,N,ACFD1,ACFD2,ACFD3,ASP,Discrete1,Discrete2,Discrete3,Discrete4,Discrete5,App2,App3,App4}
and references therein.
In comparison with the problem on the full-line or the problem on the half-line, the problem on a graph is more complicated and is not a fully developed subject yet.
However the study on this subject is currently fast growing; see for example \cite{N,ACFD1,ACFD2,ACFD3,ASP}
for recent developments regarding the integrable nonlinear Schr\"{o}dinger (NLS) equation on various simple graphs.

In this paper we show how to analyze initial-boundary value (IBV) problems for integrable nonlinear differential-difference equations (DDEs) on a graph
by using the unified transform method (UTM) \cite{F6}.
The illustrative example we choose is the Ablowitz-Ladik (AL) system \cite{AL1,AL2,AL3}:
\begin{eqnarray}
\begin{array}{l}
i\frac{d q_n}{dt}+q_{n+1}-2q_{n}+q_{n-1}-p_{n}q_{n}\left( q_{n+1}+ q_{n-1}\right)=0,
\\
i\frac{d q_n}{dt}-p_{n+1}+2p_{n}-p_{n-1}+p_{n}q_{n}\left( p_{n+1}+ p_{n-1}\right)=0,
\end{array}
 \label{aleq1}
\end{eqnarray}
where $q_n=q(n,t)$ and $p_n=p(n,t)$ are complex functions. 
We consider this integrable DDE on a graph $\mathcal{G}$ that is made of $N\geq 1$ semi-infinite lattices (edges)
connected through some boundary conditions at $n=0$ and at $n=-1$.

In order to analyze such a problem, we follow the following idea from the paper \cite{VC1} about the problem of
integrable partial differential equations (PDEs) on a star graph:
mapping the problem on a graph to a matrix IBV problem
and then extending the UTM for analyzing IBV problems in scalar case to the one in matrix case.
The present paper provides a discrete analogue of the main results of papers \cite{VC1,VC2} by Caudrelier.

We note that the UTM, introduced by Fokas \cite{F1,F6} for analyzing IBV problems of integrable PDEs,
provides an important generalization of the inverse scattering method (ISM).
The UTM has been implemented to analyze IBV problems 
for both integrable PDEs and integrable DDEs;
see for example \cite{F1,F3,F4,F5,F6,MFS1,L1,L2,FL1,XF1,XF2,GLZ} for this method on integrable PDEs
and see \cite{BH1,BH2,BH3,XFPD,Xia} for this method on integrable DDEs.

The main results derived in the present paper are stated in Proposition \ref{proposition1} and Theorem \ref{theorem1}.
Proposition \ref{proposition1} implies that the analysis of the AL lattice system (\ref{aleq1}) on the graph $\mathcal{G}$
is equivalent to the analysis of a certain matrix IBV problem.
Thus we can analyze the AL lattice system on the graph $\mathcal{G}$
by extending the UTM for integrable DDEs in the scalar case to the one in the matrix case.
Theorem \ref{theorem1} shows that the solution of the AL lattice system on the graph $\mathcal{G}$ can be expressed
in term of the solution of an appropriate matrix Riemann-Hilbert (RH) problem.
Moreover, we compare our results with some previously known studies for integrable DDEs and illustrate how our results embrace these studies as a particular case.
In particular, we show in detail that the standard ISM for the AL system on the integers (see \cite{AL3})
can be recovered as a special case of the UTM applied to our $N=2$ graph problem;
see Theorem \ref{theorem3}.
We also show that both the integrable discrete NLS (IDNLS) equation (see \cite{AL3}) and the {\it nonlocal} IDNLS equation (see \cite{AMD})
can be obtained as standard {\it local} reductions of our matrix AL system on the non-negative integers; see Proposition \ref{proposition3}.
Thus, in addition to the ISM for the IDNLS equation, the ISM for the nonlocal IDNLS equation (see \cite{AMD}) can be also recovered from our results;
see Proposition \ref{proposition4}.

The paper is organized as follows:
in Section 2, we introduce the problem of the AL system on a graph and
then we formulate this problem into a certain matrix IBV problem.
In Section 3, we implement the UTM to analyze the matrix IBV problem formulated in Section 2.
In Section 4, we compare our results with some previously known studies
and show how these previous studies can be recovered from our results.
We discuss further our results in Section 5.

\section{Problem formulation}
\subsection{AL system on a graph}
We consider the AL lattice system (\ref{aleq1}) on the graph $\mathcal{G}$.
Recall that the graph $\mathcal{G}$ is the union of $N$ edges:
each edge is made of semi-infinite lattice $\mathbb{N}_0$, the set of non-negative integers, namely $\mathbb{N}_0=\{0,1,2,\cdots\}$;
these edges are connected to each other through some boundary conditions.

We introduce $N$-copies of the AL lattice system (\ref{aleq1}) for functions $\left\{q^{\alpha}(n,t),p^{\alpha}(n,t)\right\}$, $\alpha=1,2,\cdots,N$.
The AL lattice system on the graph $\mathcal{G}$ is equivalent to a system of $N$  AL systems
such that each $\left\{q^{\alpha}(n,t),p^{\alpha}(n,t)\right\}$ lives on edge $\alpha$, is a pair of
functions of $n\in \mathbb{N}_0$ and $0< t < T$,
and the edges meet one another through some boundary conditions at $n=0$ and at $n=-1$.
Therefore the problem reads, for $\alpha=1,2,\cdots,N$,
\begin{eqnarray}
\begin{array}{l}
i\frac{dq^{\alpha}_n}{dt}+q^{\alpha}_{n+1}-2q^{\alpha}_n+q^{\alpha}_{n-1}-p^{\alpha}_{n}q^{\alpha}_{n}\left( q^{\alpha}_{n+1}+ q^{\alpha}_{n-1}\right)=0,
~~n\in \mathbb{N}_0, ~~0< t < T,
\\
i\frac{dp^{\alpha}_n}{dt}-p^{\alpha}_{n+1}+2p^{\alpha}_{n}-p^{\alpha}_{n-1}+p^{\alpha}_{n}q^{\alpha}_{n}\left( p^{\alpha}_{n+1}+ p^{\alpha}_{n-1}\right)=0,
~~n\in \mathbb{N}_0, ~~0< t < T,
\end{array}
 \label{ALN}
 \\
\begin{array}{l}
q^{\alpha}(n,0)=q_0^{\alpha}(n), ~~p^{\alpha}(n,0)=p_0^{\alpha}(n),
\\
q^{\alpha}(-1,t)=g_{-1}^{\alpha}(t), ~~q^{\alpha}(0,t)=g_{0}^{\alpha}(t), ~~p^{\alpha}(-1,t)=h_{-1}^{\alpha}(t), ~~p^{\alpha}(0,t)=h_{0}^{\alpha}(t),
\end{array}
 \label{IBVN}
\end{eqnarray}
where $q_0^{\alpha}(n)$, $p_0^{\alpha}(n)$ denote the initial data, and $g_{j}^{\alpha}(t)$, $h_{j}^{\alpha}(t)$, $j=-1,0$, denote the boundary values.
For each $\alpha$, equation (\ref{ALN}) is the discrete compatibility condition $\frac{d \mu^{\alpha}(n+1,t,z)}{dt}=\left.\frac{d \mu^{\alpha}(m,t,z)}{dt}\right|_{m=n+1}$ of
the following linear systems (called Lax pair)  \cite{Miller,XFPD}:
\begin{subequations}
\begin{eqnarray}
&&f^{\alpha}(n,t)\mu^{\alpha}(n+1,t,z)-\hat{\mathcal{Z}}\mu^{\alpha}(n,t,z)=\mathcal{U}^{\alpha}(n,t)\mu^{\alpha}(n,t,z)\mathcal{Z}^{-1},
\label{LPSN}\\
&&\mu^{\alpha}_t(n,t,z)-i\omega(z)[\sigma_3,\mu^{\alpha}(n,t,z)]=
\mathcal{V}^{\alpha}(n,t,z)\mu^{\alpha}(n,t,z),
\label{LPTN}
\end{eqnarray}
\label{LPSTN}
\end{subequations}
where $\mu^{\alpha}(n,t,z)$ is a $2\times 2$ matrix,
\begin{eqnarray}
\begin{split}
&f^{\alpha}(n,t)=\sqrt{1-q^{\alpha}(n,t)p^{\alpha}(n,t)}, ~~\omega(z)=\frac{1}{2}\left(z-z^{-1}\right)^2,
\\
&\mathcal{Z}=\left( \begin{array}{cc} z & 0 \\
 0 &  z^{-1}\\ \end{array} \right),
 \quad
\sigma_3=\left( \begin{array}{cc} 1 & 0 \\
 0 &  -1 \\ \end{array} \right),
 \\
&\mathcal{U}^{\alpha}(n,t)=\left( \begin{array}{cc} 0 & q^{\alpha}(n,t) \\
 p^{\alpha}(n,t) &  0 \\ \end{array} \right),
 \\
&\mathcal{V}^{\alpha}(n,t,z)=i\left(\mathcal{U}^{\alpha}(n-1,t)\mathcal{Z}-\mathcal{U}^{\alpha}(n,t)\mathcal{Z}^{-1}
-\frac{1}{2}\left(\mathcal{U}^{\alpha}(n,t)\mathcal{U}^{\alpha}(n-1,t)+\mathcal{U}^{\alpha}(n-1,t)\mathcal{U}^{\alpha}(n,t)\right)\right)\sigma_3,
\end{split}
\label{pqh}
\end{eqnarray}
and the symbol $\hat{\mathcal{Z}}\mu^{\alpha}(n,t,z)$ stands for $\mathcal{Z} \mu^{\alpha}(n,t,z) \mathcal{Z}^{-1}$.

\noindent
{\bf Remark 1}.
At this step we describe the problem of the AL lattice system on the graph $\mathcal{G}$
by the IBV problem (\ref{ALN}) and (\ref{IBVN}).
We should point out that we cannot view (\ref{ALN}) and (\ref{IBVN}) as $N$ disconnected copies of the semi-infinite lattice problem,
instead, we should consider (\ref{ALN}) and (\ref{IBVN}) for all $\alpha$ as a whole,
since in our context the $N$ semi-infinite lattices (edges of $\mathcal{G}$) are connected to each other through some boundary conditions.
The purpose of this section and the following section is to present the general framework for the lattice graph problem
and thus we will not focus on the connections between the edges of the graph in these two sections;
it is the purpose of Section 4 to discuss these nontrivial connections between the edges of the graph $\mathcal{G}$.

\subsection{ Mapping the problem on the graph to a certain matrix IBV problem}

We have formulated the AL lattice system (\ref{aleq1}) on the graph $\mathcal{G}$ into the IBV problem (\ref{ALN}) and (\ref{IBVN}).
We next show that analyzing such an IBV problem is equal to analyzing a certain matrix IBV problem.

We introduce the $N\times N$ diagonal-matrices
\begin{subequations}
\begin{eqnarray}
&Q(n,t)=diag \left(q^1(n,t), \cdots, q^N(n,t)\right),
\label{Q}
\\
&P(n,t)=diag \left( p^1(n,t), \cdots, p^N(n,t)\right),
\label{P}
\\
&\mathcal{F}(n,t)=diag \left(f^1(n,t),\cdots, f^N(n,t)\right).
\label{F}
\end{eqnarray}
\label{QPF}
\end{subequations}
We note that (\ref{ALN}) is equivalent to the matrix valued AL system
\begin{eqnarray}
\begin{array}{l}
i\frac{dQ_{n}}{dt}+Q_{n+1}-2Q_{n}+Q_{n-1}-P_{n}Q_{n}\left( Q_{n+1}+ Q_{n-1}\right)=0, ~~n\in \mathbb{N}_0, ~~0< t < T,
\\
i\frac{dP_{n}}{dt}-P_{n+1}+2P_{n}-P_{n-1}+P_{n}Q_{n}\left( P_{n+1}+ P_{n-1}\right)=0, ~~n\in \mathbb{N}_0, ~~0< t < T,
\end{array}
 \label{ALM}
\end{eqnarray}
where $Q_{n}=Q(n,t)$ and $P_{n}=P(n,t)$ are chosen to be the diagonal-matrices in the form of (\ref{Q}) and (\ref{P}).
Moreover, the initial and boundary conditions (\ref{IBVN}) is equivalent to
\begin{eqnarray}
\begin{split}
Q(n,0)=Q_0(n),~~P(n,0)=P_0(n),
\\
Q(-1,t)=G_{-1}(t), ~~Q(0,t)=G_0(t),
\\
P(-1,t)=H_{-1}(t), ~~P(0,t)=H_0(t),
\end{split}
\label{ibv}
\end{eqnarray}
where
\begin{eqnarray}
\begin{split}
&Q_0(n)=diag \left(q_0^{1}(n),\cdots,q_0^{N}(n)\right),
\\
&P_0(n)=diag \left(p_0^{1}(n),\cdots,p_0^{N}(n)\right),
\\
&G_l(t)=diag \left(g_l^{1}(t),\cdots,g_l^{N}(t)\right), ~~l=-1,0,
\\
&H_l(t)=diag \left(h_l^{1}(t),\cdots,h_l^{N}(t)\right), ~~l=-1,0.
\end{split}
\label{ibvdiag}
\end{eqnarray}

The matrix AL system (\ref{ALM}) is the discrete compatibility condition  of
the following linear systems
\begin{subequations}
\begin{eqnarray}
F(n,t)\mu(n+1,t,z)-\hat{Z}\mu(n,t,z)=U(n,t)\mu(n,t,z)Z^{-1},
\label{LPSM}
\\
\mu_t(n,t,z)-i\omega(z)[\Sigma_3,\mu(n,t,z)]=
V(n,t,z)\mu(n,t,z),
\label{LPTM}
\end{eqnarray}
\label{LPSTM}
\end{subequations}
where $\mu(n,t,z)$ is a $2N\times 2N$ matrix,
\begin{eqnarray}
\begin{split}
&\Sigma_3=\left( \begin{array}{cc} I_N & 0 \\
 0 &  -I_N \\ \end{array} \right),
\\
&Z=\left( \begin{array}{cc} zI_N & 0 \\
 0 &  z^{-1}I_N\\ \end{array} \right),
 \\
&F(n,t)=\left( \begin{array}{cc} \mathcal{F}(n,t) & 0 \\
 0 &  \mathcal{F}(n,t) \\ \end{array} \right),
 \\
&U(n,t)=\left( \begin{array}{cc} 0 & Q(n,t) \\
 P(n,t) &  0 \\ \end{array} \right),
 \\
&V(n,t,z)=i\left(U(n-1,t)Z-U(n,t)Z^{-1}-\frac{1}{2}\left(U(n,t)U(n-1,t)+U(n-1,t)U(n,t)\right)\right)\Sigma_3,
\end{split}
\label{FH}
\end{eqnarray}
and the symbol $\hat{Z}\mu(n,t,z)$ stands for $Z \mu(n,t,z) Z^{-1}$.

We need to introduce some notations.
Denote by $\mathbb{M}_m$ the algebra of $m\times m$ matrices over $\mathbb{C}$.
For $M\in\mathbb{M}_{2N}$, we write
$M=\left(
\begin{array}{c|c}
	M^{11}& M^{12} \\ \hline
	M^{21}& M^{22}
\end{array}
\right)$,
where the blocks $M^{jk}$, $j,k=1,2$, are $N\times N$ matrices.
We write
$M_d=\left(
\begin{array}{c|c}
	M_d^{11}& M_d^{12} \\ \hline
	M_d^{21}& M_d^{22}
\end{array}
\right)$
and
$M_o=\left(
\begin{array}{c|c}
	M_o^{11}& M_o^{12} \\ \hline
	M_o^{21}& M_o^{22}
\end{array}
\right)$,
where $M_d^{jk}$ and $M_o^{jk}$ denote the diagonal and the off-diagonal parts of $M^{jk}$, respectively.
We denote $\mathbb{M}_d =\{M_d , M \in \mathbb{M}_{2N}\}$ and $\mathbb{M}_o =\{M_o, M \in \mathbb{M}_{2N}\}$ the corresponding
sets. Consider the isomorphism
\begin{eqnarray}
\begin{split}
\theta: \qquad \prod_{j=1}^N \mathbb{M}_{2}&\rightarrow\mathbb{M}_d
\\
(M^1,\cdots,M^N)&\mapsto M=\sum_{j=1}^N M^{j}\otimes E_{jj},
\end{split}
\label{iso}
\end{eqnarray}
where $\left\{E_{jk}\right\}_{j,k=1}^N$ is the canonical basis of $\mathbb{M}_N$, and the algebra structure of $\prod_{j=1}^N \mathbb{M}_{2}$ is defined by the pointwise operations.
We find the following result \cite{VC1}:
\begin{lemma}\label{lemma1}
$\mathbb{M}_d$ and $\mathbb{M}_o$ are vector subspaces of $\mathbb{M}_{2N}$ and the direct sum decomposition  $\mathbb{M}_{2N}=\mathbb{M}_d\oplus \mathbb{M}_o$ holds.
Moreover, $\mathbb{M}_d$ is a subalgebra of $\mathbb{M}_{2N}$ which is
isomorphic to the direct product $\prod_{j=1}^N \mathbb{M}_{2}$ as algebras.
\end{lemma}

In analogy with the problem in continuum case \cite{VC1},
the key observation is that the fundamental solution of (\ref{LPSTM}) with an appropriate normalization,
is an $\mathbb{M}_d$-valued function of $n$, $t$, $z$ in the domain where it is defined.
More precisely, we have
\begin{proposition}\label{proposition1}
Let $\mu(n,t,z)$ be the fundamental solution of (\ref{LPSTM}) with normalization $\mu(n_0,t_0,z)=I_{2N}$ at a fixed point $(n_0,t_0)\in \mathbb{N}_0\times \mathbb{R}^{+}$.
Then $\mu(n,t,z)\in \mathbb{M}_d$ in the domain wherever it is defined.
\end{proposition}
{\bf Proof} This conclusion is deduced by using Lemma \ref{lemma1} and the linearity of equations (\ref{LPSTM}) for $\mu(n,t,z)$.  \QEDB

We recall that all the ingredients required for the implementation of the UTM can be derived from the fundamental solutions
of the associated Lax pair by algebraic manipulations; see for example \cite{F6,XFPD} for details.
We therefore deduce from Proposition \ref{proposition1} that the implementation of the UTM to the matrix AL system (\ref{ALM})
and thus to the AL system on the graph $\mathcal{G}$ can be entirely formulated in $\mathbb{M}_d$.

\section{Unified transformation method for the AL system on a graph}

In the above section, we have mapped the problem of the AL lattice system on the graph $\mathcal{G}$ to a $\mathbb{M}_d$-valued matrix IBV problem.
In this section, we show how to analyze such a $\mathbb{M}_d$-valued matrix IBV problem via the UTM.
The detailed derivations regarding the implementation of the UTM in the present matrix case
can be obtained easily via a similar manner as presented in the scalar case \cite{BH1,XFPD}.
For economy of presentation, here we will skip several details on these derivations
and only present the main and essential steps about the implementation of the UTM.
We refer the reader to Section 4 of \cite{XFPD} for more details on these derivations.

\subsection{ Direct part of the UTM}

\subsubsection{The eigenfunctions}
We define three different $2N\times 2N$ matrix-valued eigenfunctions $\{\mu_j(n,z,t)\}_{1}^3$
as simultaneous solutions of the linear systems (\ref{LPSTM}).
These three eigenfunctions are normalized respectively at $(n,t)=(0,0)$, at $(n,t)=(\infty,t)$, and at $(n,t)=(0,T)$.
They are given by:
\begin{eqnarray}
\begin{split}
\mu_{1}(n,t,z)=&C(n,t)C^{-1}(0,t)\left(I_{2N}+\hat{Z}^{n}\int_{0}^{t}e^{iw(z)(t-t')\hat{\Sigma}_3}\left(V\mu_{1}(0,t',z)\right)dt'\right)
\\&+C(n,t)Z^{-1}\sum_{m=0}^{n-1} C^{-1}(m,t)\hat{Z}^{n-m}(U(m,t)\mu_{1}(m,t,z)),
 \\
\mu_{2}(n,t,z)=&C(n,t)\left(I_{2N}-Z^{-1}\sum_{m=n}^{\infty} C^{-1}(m,t)\hat{Z}^{n-m}(U(m,t)\mu_{2}(m,t,z))\right),
\\
\mu_{3}(n,t,z)=&C(n,t)C^{-1}(0,t)\left(I_{2N}-\hat{Z}^{n}\int_{t}^{T}e^{iw(z)(t-t')\hat{\Sigma}_3}\left(V\mu_{3}(0,t',z)\right)dt'\right)
\\&+C(n,t)Z^{-1}\sum_{m=0}^{n-1} C^{-1}(m,t)\hat{Z}^{n-m}(U(m,t)\mu_{3}(m,t,z)),
\end{split}
\label{muibv}
\end{eqnarray}
where $I_{2N}$ denotes the $2N\times 2N$ identity matrix, and
\begin{eqnarray}
C(n,t)=\prod^{\infty}_{m=n}F(m,t),
\quad
C(-\infty)=\lim_{n\rightarrow-\infty}C(n,t),
\end{eqnarray}
and the symbol $e^{\hat{\Sigma}_3}$ acts on a $2N \times 2N$ matrix $A$ as follows
\begin{eqnarray}
e^{\hat{\Sigma}_3}A=e^{\Sigma_3}Ae^{-\Sigma_3}.
\label{ehatd}
\end{eqnarray}
Note that Proposition \ref{proposition1} implies that $\{\mu_j(n,z,t)\}_{1}^3$ are $\mathbb{M}_d$-valued functions.
Thus the spectral functions corresponding to $\{\mu_j(n,z,t)\}_{1}^3$ are also $\mathbb{M}_d$-valued functions;
this fact will become clear in Section 3.1.2.

We introduce the following domains for the AL system (see Figure 1):
\begin{subequations}
\begin{eqnarray}
&&D_{in}=\left\{z\in\mathbb{C} \Big| |z|< 1 \right\},
~~D_{out}=\left\{z\in\mathbb{C} \Big| |z|> 1 \right\},
\label{D1io}
\\
&&D_{+}=\left\{z\in\mathbb{C} \Big| \text{Im} (\omega(z))>0\right\},
~D_{-}=\left\{z\in\mathbb{C} \Big| \text{Im} (\omega(z))<0\right\},
\label{D1pm}
\\
&&D_{+in}=\left\{z\in\mathbb{C} \Big|  |z|<1, \arg z\in(\frac{\pi}{2},\pi)\cup (\frac{3\pi}{2},2\pi)\right\},
\label{D1a}
\\
&&D_{-in}=\left\{z\in\mathbb{C} \Big|  |z|<1, \arg z\in(0,\frac{\pi}{2})\cup (\pi,\frac{3\pi}{2})\right\},
\label{D1b}
\\
&&D_{+out}=\left\{z\in\mathbb{C} \Big|  |z|>1, \arg z\in(0,\frac{\pi}{2})\cup (\pi,\frac{3\pi}{2})\right\},
\label{D1c}
\\
&&D_{-out}=\left\{z\in\mathbb{C} \Big|  |z|>1, \arg z\in(\frac{\pi}{2},\pi)\cup (\frac{3\pi}{2},2\pi)\right\},
\label{D1d}
\end{eqnarray}
\label{D1}
\end{subequations}
and we denote by $\bar{D}_{in}$, $\bar{D}_{out}$, $\bar{D}_{\pm}$, $\bar{D}_{\pm in}$, $\bar{D}_{\pm out}$ the closure of these domains.
We will use the notations:
\begin{equation}
\mu_j(n,t,z)=\left(\mu_j^L(n,t,z) \mid \mu_j^R(n,t,z)\right), ~~j=1,2,3,
\end{equation}
where $\mu_j^L(n,t,z)$ is the $2N\times N$ left block and $\mu_j^R(n,t,z)$ is the $2N\times N$ right block of the matrix $\mu_j(n,t,z)$.
From (\ref{muibv}), we deduce that
\begin{itemize}
\item $\mu_{1}(n,t,z)$ is analytic for $z\in \mathbb{C}\setminus\{0\}$;  $\mu^L_{1}(n,t,z)$ is continuous and bounded for $z\in \bar{D}_{-out}$, and $\mu^R_{1}(n,t,z)$ is continuous and bounded for $z\in \bar{D}_{+in}$.
\item $\mu^L_{2}(n,t,z)$ is analytic for $D_{in}$ and it is continuous and bounded for $z\in \bar{D}_{in}$, and $\mu^R_{2}(n,t,z)$ is analytic for $D_{out}$ and it is continuous and bounded for $z\in \bar{D}_{out}$.
\item $\mu_{3}(n,t,z)$  are analytic for $z\in \mathbb{C}\setminus\{0\}$ (in the case of finite T).
If $T =\infty$, $\mu^L_{3}(n,t,z)$ is analytic for $z\in D_{+out}$, and $\mu^R_{3}(n,t,z)$ is analytic for $z\in D_{-in}$.
$\mu^L_{3}(n,t,z)$ is continuous and bounded for $z\in \bar{D}_{+out}$, and $\mu^R_{3}(n,t,z)$ is continuous and bounded for $z\in \bar{D}_{-in}$.
\end{itemize}

\begin{figure}
\begin{minipage}[t]{0.5\linewidth}
\centering
\includegraphics[width=2.5in]{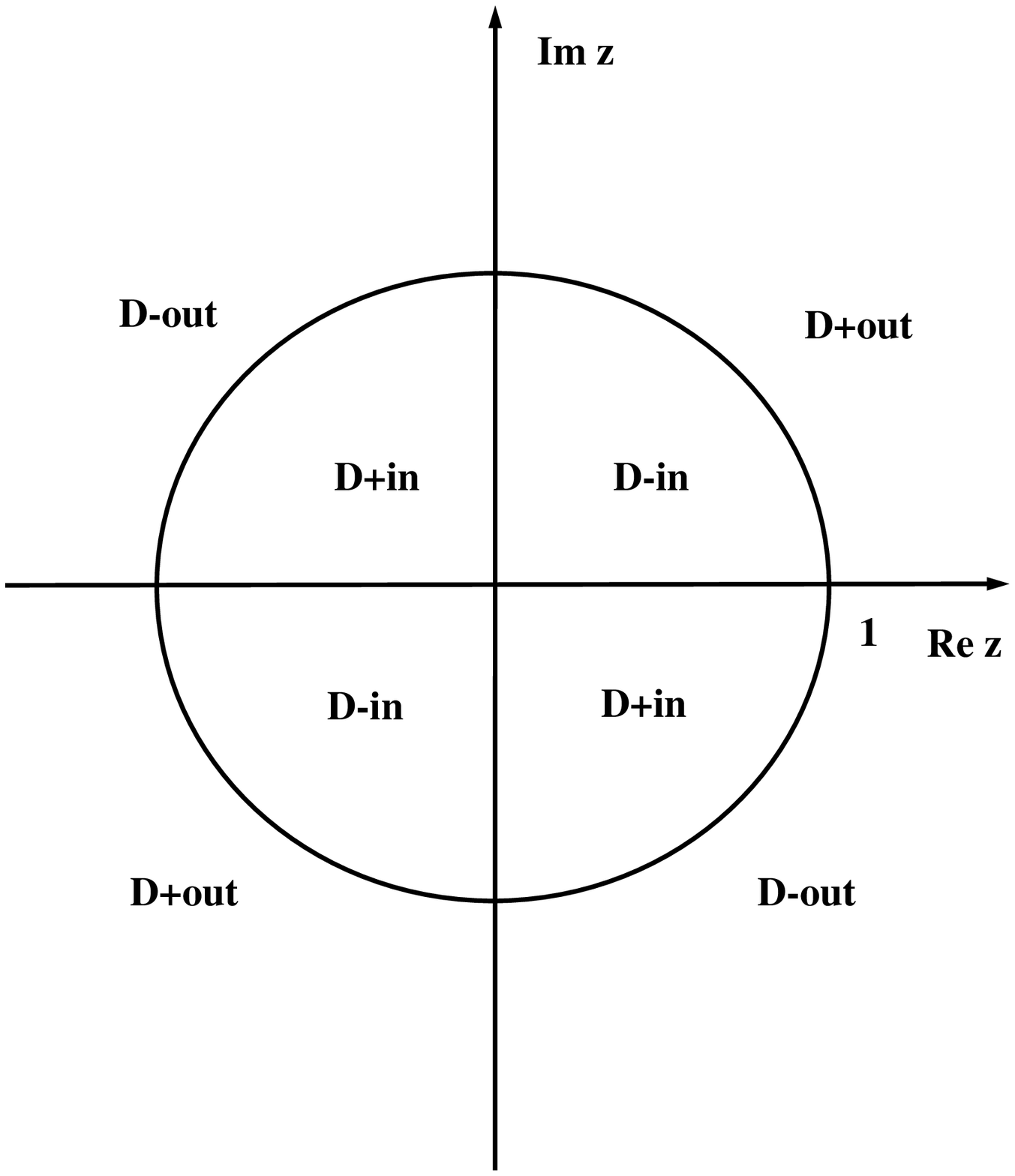}
\caption{\small{ The domains $D_{\pm in}$ and $D_{\pm out}$ of the $z$-plane for the AL equation.}}
\label{F1}
\end{minipage}
\hspace{2.0ex}
\begin{minipage}[t]{0.5\linewidth}
\centering
\includegraphics[width=2.5in]{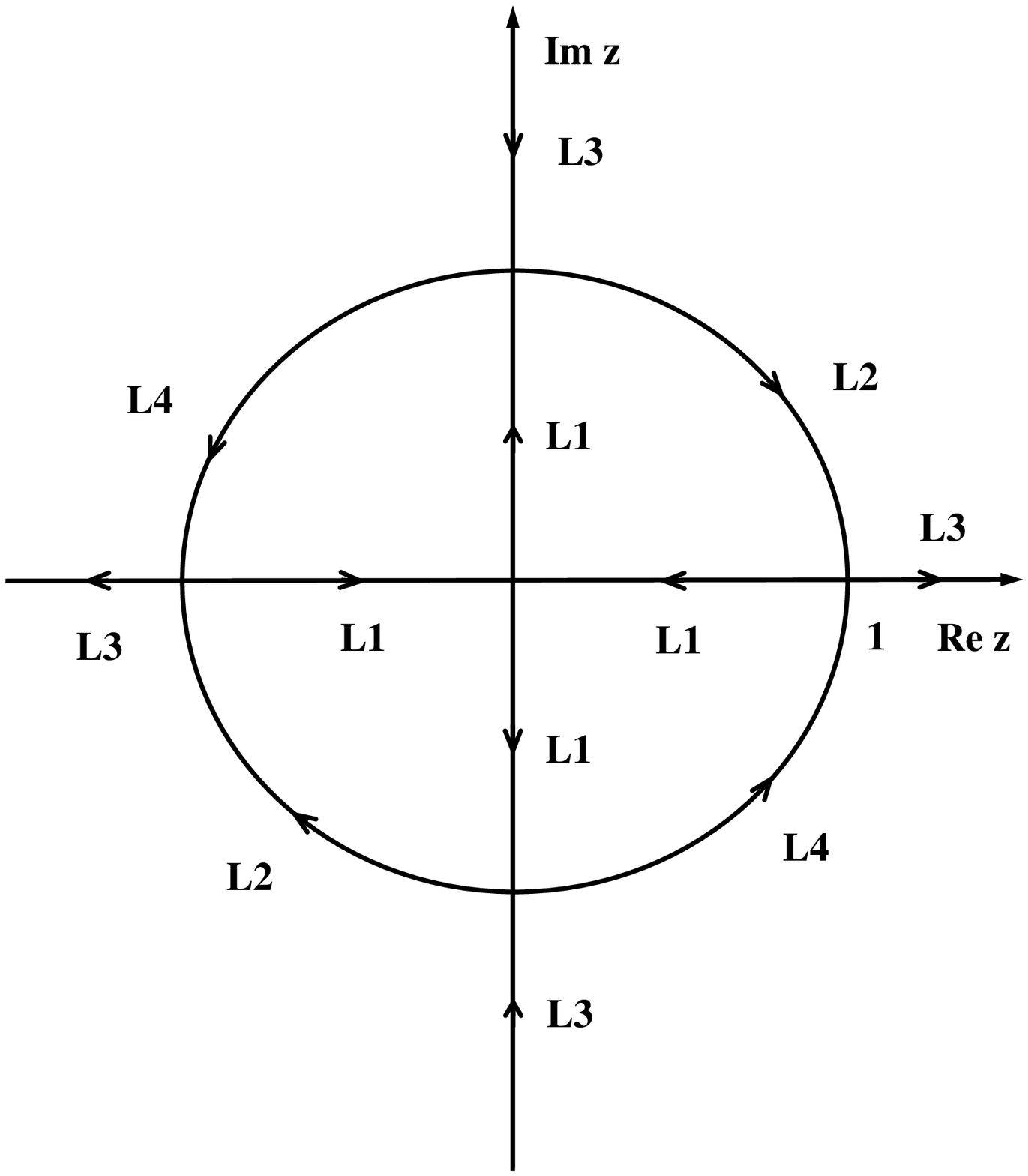}
\caption{\small{ The oriented contours $L_{1}$, $L_{2}$, $L_{3}$, $L_{4}$ for the
RH problem for the AL equation.}}
\label{F2}
\end{minipage}
\end{figure}

\subsubsection{ The spectral functions}

The matrices $\{\mu_j(n,t,z)\}_1^3$ are related:
\begin{subequations}
\begin{eqnarray}
&&\mu_2(n,t,z)=\mu_1(n,t,z)\hat{Z}^ne^{i\omega(z)t\hat{\Sigma}_3}s(z),\label{s4}
\\
&&\mu_3(n,t,z)=\mu_1(n,t,z)\hat{Z}^ne^{i\omega(z)t\hat{\Sigma}_3}S(z).
\label{s5}
\end{eqnarray}
\label{s45}
\end{subequations}
Evaluating equation (\ref{s4}) at $n=0$, $t=0$, and evaluating equation (\ref{s5}) at $n=0$, $t=T$, we obtain
\begin{subequations}
\begin{eqnarray}
&&s(z)=\mu_2(0,0,z),
\label{sSa}
\\
&&S^{-1}(z)=e^{-i\omega(z)T\hat{\Sigma}_3}\mu_1(0,T,z).
\label{sSb}
\end{eqnarray}
\label{sS}
\end{subequations}

We introduce the notations:
\begin{eqnarray}
s(z)=\left( \begin{array}{cc} \tilde{a}(z) & b(z) \\ \tilde{b}(z) & a(z) \\ \end{array} \right),
\quad
S(z)=\left( \begin{array}{cc} \tilde{A}(z) & B(z) \\ \tilde{B}(z) & A(z) \\ \end{array} \right).
\label{sm1}
\end{eqnarray}
Proposition \ref{proposition1} and the formulae (\ref{sS}) imply that
$\{a(z), b(z), \tilde{a}(z), \tilde{b}(z)\}$ and $\{A(z), B(z), \tilde{A}(z), \tilde{B}(z)\}$
are all $N \times N$ diagonal-matrix valued functions (called spectral functions),
and they have the following properties:
\begin{itemize}
\item the entries of $a(z)$, $b(z)$ are analytic for $|z| > 1$ and continuous and bounded for $|z| \geq 1$.
\item the entries of $\tilde{a}(z)$, $\tilde{b}(z)$ are analytic for $|z| < 1$ and continuous and bounded for $|z| \leq 1$.
\item the entries of $A(z)$, $B(z)$ are analytic for $z\in\mathbb{C}\setminus\{0\}$ and continuous and bounded for $z\in\bar{D}_-$ (in the case of finite T).
If $T =\infty$, the functions $A(z)$ and $B(z)$ are defined and analytic for $z\in D_-$.
\item the entries of $\tilde{A}(z)$, $\tilde{B}(z)$ are analytic for $z\in\mathbb{C}\setminus\{0\}$ and continuous and bounded for $z\in\bar{D}_+$ (in the case of finite T).
    If $T =\infty$, the functions $\tilde{A}(z)$ and $\tilde{B}(z)$ are defined and analytic for $z\in D_+$.
\end{itemize}

The Lax pair (\ref{LPSTM}) implies that
\begin{subequations}
\begin{eqnarray}
a(z)\tilde{a}(z)-b(z)\tilde{b}(z)=I_N,
\\
A(z)\tilde{A}(z)-B(z)\tilde{B}(z)=I_N.
\end{eqnarray}
\label{det2}
\end{subequations}
These identities immediately yield
\begin{eqnarray}
s^{-1}(z)=\left( \begin{array}{cc} a(z) & -b(z) \\ -\tilde{b}(z) & \tilde{a}(z) \\ \end{array} \right),
~~
S^{-1}(z)=\left( \begin{array}{cc} A(z) & -B(z) \\ -\tilde{B}(z) & \tilde{A}(z) \\ \end{array} \right).
\label{sm2}
\end{eqnarray}

\subsubsection{The symmetry properties}

In analogy to the scalar case \cite{BH1,XFPD}, from equations (\ref{muibv}), (\ref{sS}) and (\ref{sm1}), we find the following symmetry relations regarding the spectral functions:
\begin{eqnarray}
\begin{split}
a(-z)&=a(z),\quad b(-z)=-b(z), \quad \tilde{a}(-z)=\tilde{a}(z),\quad \tilde{b}(-z)=-\tilde{b}(z),
\\
A(-z)&=A(z),\quad B(-z)=-B(z), \quad \tilde{A}(-z)=\tilde{A}(z),\quad \tilde{B}(-z)=-\tilde{B}(z).
\end{split}
\label{sr}
\end{eqnarray}
Let us introduce the following functions:
\begin{eqnarray}
\begin{split}
\gamma(z)&=b(z)\tilde{a}^{-1}(z), ~~\tilde{\gamma}(z)=\tilde{b}(z)a^{-1}(z),
\\
R(z)&=B(z)A^{-1}(z),~~ \tilde{R}(z)=\tilde{B}(z)\tilde{A}^{-1}(z),
\\
d(z)&=\tilde{a}(z)A(z)-\tilde{b}(z)B(z),
~~
\tilde{d}(z)=a(z)\tilde{A}(z)-b(z)\tilde{B}(z),
\\
\Gamma(z)&=\tilde{B}(z)a^{-1}(z)\tilde{d}^{-1}(z), ~~\tilde{\Gamma}(z)=B(z)\tilde{a}^{-1}(z)d^{-1}(z).
\end{split}
\label{grd}
\end{eqnarray}
It follows from (\ref{sr}) that
\begin{eqnarray}
\begin{split}
\gamma(-z)&=-\gamma(z),\quad \tilde{\gamma}(-z)=-\tilde{\gamma}(z),
\\
d(-z)&=d(z),\quad \tilde{d}(-z)=\tilde{d}(z),
\\
\Gamma(-z)&=-\Gamma(z),\quad \tilde{\Gamma}(-z)=-\tilde{\Gamma}(z).
\end{split}
\label{sr1}
\end{eqnarray}

\subsubsection{A Riemann-Hilbert problem}
We define $M(n,t,z)=M_{+}(n,t,z)$, for $z\in \bar{D}_{+}$, and $M(n,t,z)=M_{-}(n,t,z)$, for $z\in \bar{D}_{-}$, where $M_{\pm}(n,t,z)$ are defined by
\begin{eqnarray}
\begin{split}
M_{+}(n,t,z)&=\left\{\begin{array}{l}C^{-1}(n,t)\left(\mu_2^L(n,t,z),\left(I_2\otimes \tilde{a}(z)\right)^{-1}\mu_1^R(n,t,z)\right), \quad z\in \bar{D}_{+in},
\\
C^{-1}(n,t)\left(\left(I_2\otimes \tilde{d}(z)\right)^{-1}\mu_3^L(n,t,z),\mu_2^R(n,t,z)\right), \quad z\in \bar{D}_{+out},
\end{array}\right.
 \\
M_{-}(n,t,z)&=\left\{\begin{array}{l}C^{-1}(n,t)\left(\mu_2^L(n,t,z),\left(I_2\otimes d(z)\right)^{-1}\mu_3^R(n,t,z)\right), \quad z\in \bar{D}_{-in},
\\
C^{-1}(n,t)\left(\left(I_2\otimes a(z)\right)^{-1}\mu_1^L(n,t,z),\mu_2^R(n,t,z)\right), \quad z\in \bar{D}_{-out}.
\end{array}\right.
\end{split}
\label{M1}
\end{eqnarray}
From (\ref{muibv}) and (\ref{M1}) one can deduce that 
\begin{eqnarray}
\begin{split}
M(n,t,z)=I_{2N}+U(n-1,t)Z^{-1}+\left( \begin{array}{cc} O_{N \times N}(z^{-2},\text{even}) & O_{N \times N}(z^{3},\text{odd})  \\
  O_{N \times N}(z^{-3},\text{odd}) & O_{N \times N}(z^{2},\text{even}) \\ \end{array} \right), ~ z\rightarrow (\infty,0).
\end{split}
\label{Masp}
\end{eqnarray}
Here the notation $O_{N \times N}(z^{2},even)$ ($O_{N \times N}(z^{-2},even)$) indicates that the remaining terms are $N \times N$ diagonal-matrices
with even powers of $z$ ($z^{-1}$) in diagonal-elements;
the notation $O_{N \times N}(z^{3},\text{odd})$ ($O_{N \times N}(z^{-3},\text{odd})$) indicates that the remaining terms are $N \times N$ diagonal-matrices
with odd powers of $z$ ($z^{-1}$) in diagonal-elements;
the symbol $z\rightarrow (\infty,0)$ means $z\rightarrow \infty$ for the $2N\times N$ left block,
while $z\rightarrow 0$ for the $2N\times N$ right block of the $2N\times 2N$ matrix $M(n,t,z)$.

Using (\ref{s45}), we can formulate (\ref{M1}) into the following $2N \times 2N$ matrix RH problem
\begin{eqnarray}
M_{-}(n,t,z)=M_{+}(n,t,z)J(n,t,z), \quad z\in L=L_1\cup L_2\cup L_3\cup L_4,
\label{RHP}
\end{eqnarray}
where $J(n,t,z)=J_j(n,t,z)$ for $z\in L_j$, $j=1,2,3,4$, and these jump matrices are defined by
\begin{eqnarray}
\begin{split}
J_1(n,t,z)&=\hat{Z}^ne^{i\omega(z)t\hat{\Sigma}_3}\left( \begin{array}{cc} I_{N} & \tilde{\Gamma}(z) \\
  0 & I_{N} \\ \end{array} \right), \quad z\in L_1,
\\
J_3(n,t,z)&=\hat{Z}^ne^{i\omega(z)t\hat{\Sigma}_3}\left( \begin{array}{cc} I_{N} & 0 \\
  -\Gamma(z) & I_{N} \\ \end{array} \right), \quad z\in L_3,
\\
J_4(n,t,z)&=\hat{Z}^ne^{i\omega(z)t\hat{\Sigma}_3}\left( \begin{array}{cc} I_{N}-\gamma(z)\tilde{\gamma}(z) & \gamma(z) \\
  -\tilde{\gamma}(z) & I_{N} \\ \end{array} \right), \quad z\in L_4,
\\
J_2(n,t,z)&=J_3(n,t,z)(J_4(n,t,z))^{-1}J_1(n,t,z), \quad z\in L_2,
\end{split}
\label{JM}
\end{eqnarray}
and the contours $L_j$, $j=1,2,3,4$, are defined by (see Figure \ref{F2}):
\begin{eqnarray}
L_1=\bar{D}_{-in}\cap \bar{D}_{+in},~~ L_2=\bar{D}_{-in}\cap \bar{D}_{+out},
~~ L_3=\bar{D}_{-out}\cap \bar{D}_{+out},~~ L_4=\bar{D}_{-out}\cap \bar{D}_{+in},
\label{L}
\end{eqnarray}

The blocks $(12)$ and $(21)$ of (\ref{Masp}) induce the following expressions:
\begin{eqnarray}
\begin{split}
Q(n,t)=\lim_{z\rightarrow 0}(z^{-1}M(n+1,t,z))^{12},
\\
P(n,t)=\lim_{z\rightarrow \infty}(zM(n+1,t,z))^{21},
\end{split}
\label{solution1}
\end{eqnarray}
where the indexes ``12" and ``21" denote the blocks $(12)$ and $(21)$ in the natural decomposition of a matrix in $\mathbb{M}_d$.

If the spectral functions $a(z)$, $\tilde{a}(z)$, $d(z)$ and $\tilde{d}(z)$ can have zeros,
then $M(n,t,z)$ is a meromorphic function of $z$.
In this case, the RH problem (\ref{RHP}) is singular, thus, one has to consider the corresponding residue relations.
For details on these zeros and residue relations for the AL system in scalar case, one can refer to \cite{BH1,XFPD},
in which the same derivation goes over to the matrix case in the present paper.
For economy of presentation, here we will not consider the possibility of such zeros for spectral functions
since this respect is not essential for the main results of the present paper.

\subsubsection{The Global relation}

Using (\ref{sSb}) in (\ref{s4}) with $n=0$, $t =T$, we obtain
\begin{eqnarray}
\mu_2(0,T,z)=e^{i\omega(z)T\hat{\Sigma}_3}\left(S^{-1}(z)s(z)\right).
\label{gr1}
\end{eqnarray}
Using the definition of $\mu_2(n,t,z)$ (the second equation of (\ref{muibv})) in the above formula, we obtain
\begin{eqnarray}
S^{-1}(z)s(z)=C(0,T)-e^{-i\omega(z)T\hat{\Sigma}_3}G(z,T),
\label{gr2}
\end{eqnarray}
where
\begin{eqnarray}
G(z,t)=C(0,t)Z^{-1}\sum_{m=0}^{\infty} C^{-1}(m,t)\hat{Z}^{-m}\left(U(m,t)\mu_{2}(m,t,z)\right).
\label{gr3}
\end{eqnarray}
The $(12)$-block and $(21)$-block of equation (\ref{gr2}) yield the following global relation
\begin{subequations}
\begin{eqnarray}
&A(z)b(z)-B(z)a(z)=-e^{-2i\omega(z)T}G^{12}(z,T), \quad |z|>1,
\label{gra}
\\
&\tilde{A}(z)\tilde{b}(z)-\tilde{B}(z)\tilde{a}(z)=-e^{2i\omega(z)T}G^{21}(z,T), \quad |z|<1,
\label{grb}
\end{eqnarray}
\label{gr}
\end{subequations}
where $G^{12}(z,T)$ and $G^{21}(z,T)$ are $(12)$-block and $(21)$-block of the matrix $G(z,T)$.
As $T= \infty$, the global relation (\ref{gr}) becomes
\begin{eqnarray}
\begin{split}
A(z)b(z)-B(z)a(z)=0, \quad z\in \bar{D}_{-out},
\\
\tilde{A}(z)\tilde{b}(z)-\tilde{B}(z)\tilde{a}(z)=0, \quad z\in \bar{D}_{+in}.
\end{split}
\label{gr5}
\end{eqnarray}

\subsection{Inverse part of the UTM}

Consider the initial data $Q_0(n)$ and  $P_0(n)$, and the boundary values $G_l(t)$ and  $H_l(t)$, $l=-1,0$; see (\ref{ibvdiag}).
Denote by $l^1(\mathbb{N}_0)$ the space of sequences $\left\{a_n\right\}_{n\in \mathbb{N}_0}$, such that $\sum_{n=0}^{n=+\infty}|a_n|<\infty$.
Denote by $U_0(n)$ the matrix $U(n,0)$, in which $Q(n, 0)$ and  $P(n, 0)$ are replaced respectively by $Q_0(n)$ and  $P_0(n)$.
Denote by $F_0(n)$ the matrix $F(n,0)$, in which $\mathcal{F}(n,0)$ is replaced by $\mathcal{F}_0(n)=diag \left(\sqrt{1-q_0^{1}(n)p_0^{1}(n)},\cdots,\sqrt{1-q_0^{N}(n)p_0^{N}(n)}\right)$.
Denote by $W_l(t)$ the matrix $U(l,t)$, $l=-1,0$, in which $Q(l, t)$ and  $P(l, t)$ are replaced respectively by $G_l(t)$ and  $H_l(t)$;
denote
\begin{eqnarray}
\begin{split}
V_0(t,z)=i\left(W_{-1}(t)Z-W_{0}(t)Z^{-1}-\frac{1}{2}\left(W_{0}(t)W_{-1}(t)+W_{-1}(t)W_{0}(t)\right)\right)\Sigma_3.
\end{split}
\label{V0t}
\end{eqnarray}

Motivated by Section 3.1, we define the spectral functions as follows.
\begin{definition} 
Given $Q_0(n)$, $P_0(n)$ such that the entries of them belong to $l^1(\mathbb{N}_0)$,
define the $2N\times 2N$ matrix valued function $\phi(n, z)$
by the unique solution of
\begin{eqnarray}
\begin{split}
&F_0(n)\phi(n+1,z)-\hat{Z}\phi(n,z)=U_0(n)\phi(n,z)Z^{-1},
\\
&\lim_{n\rightarrow \infty}\phi(n, z)=I_{2N}.
\end{split}
\label{def1phi}
\end{eqnarray}
Furthermore, we define the $N \times N$ diagonal-matrix valued spectral functions
$a(z)$, $b(z)$, $\tilde{a}(z)$ and $\tilde{b}(z)$ by
\begin{eqnarray}
\begin{split}
a(z)=\phi^{22}(0,z),~~b(z)=\phi^{12}(0,z),  ~~|z|\geq 1,
\\
\tilde{a}(z)=\phi^{11}(0,z),~~\tilde{b}(z)=\phi^{21}(0,z),  ~~|z|\leq 1,
\end{split}
\label{dab}
\end{eqnarray}
where the indexes ``11", ``12", ``21" and ``22" denote respectively the blocks $(11)$, $(12)$, $(21)$ and $(22)$ in the natural decomposition of $2N\times 2N$ matrices in $\mathbb{M}_d$.
\end{definition}

\begin{definition} 
Given $G_l(t)$, $H_l(t)$, $l=-1,0$, such that the entries of them are smooth functions for $0 < t < T$,
define the $2N\times 2N$ matrix valued function $\varphi(t, z)$ by the unique solution of
\begin{eqnarray}
\begin{split}
&\varphi_t(t,z)-i\omega(z)[\Sigma_3,\varphi(t,z)]=
V_0(t,z)\varphi(t,z),
\\
&\varphi(0, z) =I_{2N}.
\end{split}
\label{ABphidef}
\end{eqnarray}
Furthermore, we define the $N \times N$ diagonal-matrix valued spectral functions
 $A(z)$, $B(z)$, $\tilde{A}(z)$ and $\tilde{B}(z)$ by
\begin{eqnarray}
\begin{split}
&A(z)=\varphi^{11}(T,z), ~~B(z)=-e^{-2i\omega(z)T}\varphi^{12}(T,z), ~~z\in \mathbb{C}\setminus \{0\},
\\
&\tilde{A}(z)=\varphi^{22}(T,z), ~~\tilde{B}(z)=-e^{2i\omega(z)T}\varphi^{21}(T,z), ~~z\in \mathbb{C}\setminus \{0\},
\end{split}
\label{dAB1}
\end{eqnarray}
where, as before, the indexes ``11", ``12", ``21" and ``22" denote respectively the blocks $(11)$, $(12)$, $(21)$ and $(22)$ in the natural decomposition of $2N\times 2N$ matrices in $\mathbb{M}_d$.
\end{definition}

If $T=\infty$, we assume that the entries of $G_l(t)$, $H_l(t)$, $l=-1,0$, belong to the Schwartz class,
and we use an alternative definition of the spectral functions $\{A(z), B(z), \tilde{A}(z), \tilde{B}(z)\}$
based on the solution $\mu_3(0, t, z)$;
namely we let
\begin{eqnarray}
\begin{split}
&A(z)=\eta^{22}(0,z), ~~B(z)=\eta^{12}(0,z), ~~\text{Im} \left(\omega(z)\right)\leq 0,
\\
&\tilde{A}(z)=\eta^{11}(0,z), ~~\tilde{B}(z)=\eta^{21}(0,z), ~~\text{Im} \left(\omega(z)\right)\geq 0,
\end{split}
\label{dAB2}
\end{eqnarray}
where $\eta(t,z)=\left( \begin{array}{cc} \eta^{11}(t,z) & \eta^{12}(t,z) \\ \eta^{21}(t,z) & \eta^{22}(t,z) \\ \end{array} \right)$ is the unique solution of
\begin{eqnarray}
\begin{split}
&\eta_t(t,z)-i\omega(z)[\Sigma_3,\eta(t,z)]=
V_0(t,z)\eta(t,z),
\\
&\lim_{t\rightarrow\infty}\eta(t, z) =I_{2N}.
\end{split}
\label{ABphidef2}
\end{eqnarray}

The main result for the AL equation on the graph $\mathcal{G}$ is the following:
\begin{theorem}\label{theorem1}
Let spectral functions $\{a(z),b(z),\tilde{a}(z),\tilde{b}(z)\}$, corresponding to $\left\{Q_0(n), P_0(n)\right\}$, be defined according to Definition 1.
Suppose that there exist $\left\{G_l(t), H_l(t)\right\}$, $l=-1,0$, such that the spectral functions $\{A(z),B(z),\tilde{A}(z),\tilde{B}(z)\}$
defined by Definition 2, satisfy the global relation (\ref{gr}).
Define $M(n, t, z)$ as the solution of the following $2N \times 2N$ matrix RH problem\footnote{Recall that we do not consider, in the present paper,
the possibility of the poles of $M(n, t, z)$ for conciseness.
}:
\begin{itemize}
\item $M(n, t, z)$ is analytic for $z\in \mathbb{C}\backslash L$, where $L=L_1\cup L_2\cup L_3\cup L_4$, and $\{L_j\}_{1}^{4}$ are defined by (\ref{L});
see Figure 2 for these contours.
\item \begin{eqnarray}
M_{-}(n,t,z)=M_{+}(n,t,z)J(n,t,z), \quad z\in L,
\label{RHP3}
\end{eqnarray}
where $M(n,t,z)=M_{\pm}(n,t,z)$ for $z\in \bar{D}_{\pm}$, $J(n,t,z)=J_{j}(n,t,z)$ for $z\in L_j$, $j=1,2,3,4$,
and the jump matrices $J_j(n,t,z)$ are defined in terms of spectral functions by (\ref{JM});
see Figure 1 and Figure 2 for these domains and contours.
\item \begin{eqnarray}
\begin{split}
&M(n,t,z)=I_{2N}+\left( \begin{array}{cc} O_{N\times N}(z^{-2},\text{even}) & O_{N\times N}(z,\text{odd})  \\
  O_{N\times N}(z^{-1},\text{odd}) & O_{N\times N}(z^2,\text{even}) \\ \end{array} \right), \quad z\rightarrow (\infty,0).
\end{split}
\label{Masp1Th}
\end{eqnarray}
\end{itemize}
Then $M(n,t, z)$ exists and is unique.
Furthermore, let
\begin{eqnarray}
Q(n,t)=\lim_{z\rightarrow 0}(z^{-1}M(n+1,t,z))^{12}, \quad P(n,t)=\lim_{z\rightarrow \infty}(zM(n+1,t,z))^{21}.
\label{solution}
\end{eqnarray}
then $Q(n,t)$, $P(n,t)$ solves the matrix AL system (\ref{ALM}) as well as satisfies initial-boundary value conditions (\ref{ibv}).
\end{theorem}
{\bf Proof} \quad
Similarly with the problem in the continuum case \cite{VC1},
we transform the proof of this theorem in the matrix case to the proof of
$N$-copies of the analogous theorem in the scalar case.
We consider the isomorphism $\theta$ defined by (\ref{iso}).
For this $\theta$, we denote by $\left(M^1(n,t,z), \cdots, M^N(n,t,z)\right)$ the preimage of $M(n,t,z)$.
Then each $M^\alpha(n,t,z)$ is defined as the solution of the $2\times2$ RH problem that takes the same form as the above one
but formulated by the spectral functions $\{a^\alpha(z)$,
$b^\alpha(z)$, $\tilde{a}^\alpha(z)$, $\tilde{b}^\alpha(z)\}$ corresponding to $\{q_0^{\alpha}(n)$, $p_0^{\alpha}(n)\}$,
and by the spectral functions $\{A^\alpha(z)$, $B^\alpha(z)$, $\tilde{A}^\alpha(z)$, $\tilde{B}^\alpha(z)\}$
corresponding to $\{g_{-1}^{\alpha}(t), ~~g_{0}^{\alpha}(t), ~~h_{-1}^{\alpha}(t), ~~h_{0}^{\alpha}(t)\}$.
Note that the formula (\ref{solution}) is equivalent to setting
$q^{\alpha}(n,t)=\lim_{z\rightarrow 0}(z^{-1}M^{\alpha}(n+1,t,z))^{12}$, $p^{\alpha}(n,t)=\lim_{z\rightarrow \infty}(zM^{\alpha}(n+1,t,z))^{21}$ for $\alpha=1,\cdots,N$.
For the latter scalar case, it follows from \cite{XFPD} that each $M^{\alpha}(n,t,z)$ exists uniquely,
and $q^{\alpha}(n,t)$, $p^{\alpha}(n,t)$ solves the AL equation with the initial-boundary conditions:
$q^{\alpha}(n,0)=q_0^{\alpha}(n)$, $p^{\alpha}(n,0)=p_0^{\alpha}(n)$, $q^{\alpha}(-1,t)=g_{-1}^{\alpha}(t), ~~q^{\alpha}(0,t)=g_{0}^{\alpha}(t), ~~p^{\alpha}(-1,t)=h_{-1}^{\alpha}(t), ~~p^{\alpha}(0,t)=h_{0}^{\alpha}(t)$.
Hence, $Q(n,t)$, $P(n,t)$ defined by (\ref{solution}), solves the AL equation (\ref{ALM}) and satisfies the initial-boundary conditions (\ref{ibv}).
\QEDB

\subsection {Analysis of the global relation}
For the matrix version of AL equation (\ref{ALM}), the matrix-valued unknown boundary values $G_0(t)$, $H_0(t)$ can be characterized as follows.

\begin{proposition}
Denote by $\partial D_{\pm in}$ and $\partial D_{\pm out}$ the oriented boundaries of $D_{\pm in}$ and $D_{\pm out}$,
such that $D_{\pm in}$ and $D_{\pm out}$ lie in the left hand side of the increasing direction.
Let $\varphi(t,z)$ be defined by (\ref{ABphidef});
let $\varphi^{11}(t,z)$, $\varphi^{12}(t,z)$, $\varphi^{21}(t,z)$, $\varphi^{22}(t,z)$
be the blocks $(11)$, $(12)$, $(21)$, $(22)$ in the natural decomposition of $\varphi(t,z)$.
The unknown boundary values $G_0(t)$, $H_0(t)$ associated with the AL equation (\ref{aleq1}) are given by
\begin{eqnarray}
\begin{split}
&G_0(t)+G_{-1}(t)=\frac{1}{\pi i}\left[\int_{\partial D_{+in}}\psi (t,z)dz+\mathcal{E}_1(t)\int_{\partial D_{+out}}
e^{2i\omega(z)t}b(z)a^{-1}(z)\varphi^{11} (t,z)dz\right],
\\
&H_0(t)+H_{-1}(t)=\frac{1}{\pi i}\left[\int_{\partial D_{-in}}\tilde{\psi} (t,z)dz+\mathcal{E}_1(t)\int_{\partial D_{-out}}
e^{-2i\omega(z)t}\tilde{b}(z^{-1})\tilde{a}^{-1}(z^{-1})\varphi^{22} (t,z^{-1})dz\right],
\end{split}
\label{DN1}
\end{eqnarray}
where
\begin{eqnarray}
\begin{split}
&\psi(t,z)=z^{-2}\left(\mathcal{E}^{-1}_1(t)\varphi^{12}(t,z)-\mathcal{E}_1(t)\varphi^{12}(t,z^{-1})\right),
\\
&\tilde{\psi}(t,z)=z^{-2}\left(\mathcal{E}^{-1}_1(t)\varphi^{21}(t,z^{-1})-\mathcal{E}_1(t)\varphi^{21}(t,z)\right),
\\
&\mathcal{E}_1(t)=diag\left(e^{\frac{i}{2}\int_{0}^t\left( g^1_0(\xi)h^1_{-1}(\xi)-g^1_{-1}(\xi)h^1_{0}(\xi)\right)d\xi},
\cdots,e^{\frac{i}{2}\int_{0}^t\left( g^N_0(\xi)h^N_{-1}(\xi)-g^N_{-1}(\xi)h^N_{0}(\xi)\right)d\xi}\right).
\end{split}
\label{psinls}
\end{eqnarray}
\end{proposition}
{\bf Proof} \quad By employing Lemma \ref{lemma1} and Proposition \ref{proposition1}, we can map the proof of the formulae (\ref{DN1}) for $G_0(t)$, $H_0(t)$ in matrix version to
the proof of the analogous formulae for each $g_{0}^{\alpha}(t)$, $h_{0}^{\alpha}(t)$, $\alpha=1,\cdots,N$, in scalar version.
For $\alpha=1,\cdots,N$, the analogous formulae for each $g_{0}^{\alpha}(t)$, $h_{0}^{\alpha}(t)$ in scalar version can be derived by the analysis of the global relation and
by the use of certain asymptotic considerations of the eigenfunctions; for details see Section 5.2 of \cite{XFPD}.
\QEDB

\section{Comparison with Previous Results}

In the above two sections, by an algebraic formulation and then by implementing the UTM to the resulting matrix IBV problem,
we have presented a framework to analyze the problem of the AL system (\ref{aleq1}) on the graph $\mathcal{G}$.
For the scalar case $N=1$, we immediately recover the results for the problem of AL system on the set of non-negative integers \cite{XFPD}.
For $N\geq 2$, as mentioned earlier, for a genuine graph, the $N$ edges are not separated,
instead, they are connected to each other through appropriate boundary conditions.
It is the goal of this section to discuss these nontrivial connections and
show how some previously known studies fit within our framework.

\subsection{Recovering the problem on the integers}

In this subsection, we show that the standard ISM for the AL system on the set of integers
can be recovered as a special case of UTM applied to our $N=2$ graph problem
with the two edges of the graph connected through the following boundary condition 
\begin{eqnarray}
\begin{split}
&G_0(t)=\sigma G_{-1}(t)\sigma,
\\
&H_0(t)=\sigma H_{-1}(t)\sigma,
\end{split}
\label{bvint}
\end{eqnarray}
where
\begin{eqnarray}
\sigma=\left( \begin{array}{cc} 0 & 1 \\
 1 &  0\\ \end{array} \right),~~G_l(t)=\left( \begin{array}{cc} g_l^{1}(t) & 0 \\
 0 &  g_l^{2}(t)\\ \end{array} \right),~~H_l(t)=\left( \begin{array}{cc} h_l^{1}(t) & 0 \\
 0 &  h_l^{2}(t)\\ \end{array} \right), ~~l=-1,0.
\label{sigma}
\end{eqnarray}

\subsubsection{ISM on the integers}

In order to compare the ISM on the integers with our $N=2$ matrix UTM on the non-negative integers,
it is very convenience to put the two problems in the same size.
More precisely, we will use a redundant $4\times 4$ Lax pair formulation instead of the standard $2\times 2$ one for the implementation of ISM for the AL equation.

We consider the following $4 \times 4$ Lax pair formulation
\begin{subequations}
\begin{eqnarray}
&&F^{int}(n,t)\mu^{int}(n+1,t,z)-\hat{Z}\mu^{int}(n,t,z)=U^{int}(n,t)\mu^{int}(n,t,z)Z^{-1},
\label{LPSint}\\
&&\mu^{int}_t(n,t,z)-i\omega(z)[\Sigma_3,\mu^{int}(n,t,z)]=V^{int}(n,t,z)\mu^{int}(n,t,z),
\label{LPTint}
\end{eqnarray}
\label{LPSTint}
\end{subequations}
where $\mu^{int}(n,t,z)$ is a $4 \times 4$ matrix, and
\begin{eqnarray}
\begin{split}
&F^{int}(n,t)=diag\left(f(n,t),f(-n-1,t),f(n,t),f(-n-1,t)\right),
\\
&Z=\left( \begin{array}{cc} zI_2 & 0 \\
 0 &  z^{-1}I_2\\ \end{array} \right),
 ~~\Sigma_3=\left( \begin{array}{cc} I_2 & 0 \\
 0 &  -I_2 \\ \end{array} \right),
 \\
&U^{int}(n,t)=\left( \begin{array}{cc} 0 & Q^{int}(n,t) \\
 P^{int}(n,t) &  0 \\ \end{array} \right),
 \\
&V^{int}(n,t,z)=i\Big(U^{int}(n-1,t)Z-U^{int}(n,t)Z^{-1}-\frac{1}{2}U^{int}(n,t)U^{int}(n-1,t)
\\&~~~~~~~~~~~~~~~~~-\frac{1}{2}U^{int}(n-1,t)U^{int}(n,t)\Big)\Sigma_3,
\end{split}
\label{FHint}
\end{eqnarray}
with
\begin{eqnarray}
\begin{split}
&f(n,t)=\sqrt{1-q(n,t)p(n,t)},
\\
&Q^{int}(n,t)=\left( \begin{array}{cc} q(n,t) & 0 \\
 0 &  q(-n-1,t)\\ \end{array} \right),
 \\
&P^{int}(n,t)=\left( \begin{array}{cc} p(n,t) & 0 \\
 0 &  p(-n-1,t)\\ \end{array} \right).
\end{split}
\label{fQPint}
\end{eqnarray}
The compatibility condition of (\ref{LPSTint}) gives rise to the following $2\times 2$ matrix AL equation
\begin{eqnarray}
\begin{split}
i\frac{d Q^{int}_n}{dt}+Q^{int}_{n+1}-2Q^{int}_{n}+Q^{int}_{n-1}-P^{int}_{n}Q^{int}_{n}\left( Q^{int}_{n+1}+ Q^{int}_{n-1}\right)=0,
\\
i\frac{d P^{int}_n}{dt}-P^{int}_{n+1}+2P^{int}_{n}-P^{int}_{n-1}+P^{int}_{n}Q^{int}_{n}\left( P^{int}_{n+1}+ P^{int}_{n-1}\right)=0,
\end{split}
 \label{ALint}
\end{eqnarray}
where $Q^{int}_n=Q^{int}(n,t)$ and $P^{int}_n=P^{int}(n,t)$.
It is clear that reconstructing $q(n,t)$, $p(n,t)$ is equivalent to reconstructing $Q^{int}(n,t)$, $P^{int}(n,t)$.
The standard ISM for AL equation in the case of $2\times 2$ Lax pair formulation \cite{AL3,XFPD}
can be easily generalized to the above (redundant) $4\times 4$ case.
Here we only present the essential results; for details see \cite{AL3,XFPD}.

We need to define two particular solutions of equation (\ref{LPSint}) with normalizations as $n\rightarrow \mp \infty$.
They are given by the following summation equations:
\begin{subequations}
\begin{flalign}
\mu^{int}_{-}(n,t,z)=C^{int}(n,t)\bigg(\left(C^{int}(-\infty)\right)^{-1}+Z^{-1}\sum_{m=-\infty}^{n-1} \left(C^{int}(m,t)\right)^{-1}\hat{Z}^{n-m}(Q^{int}(m,t)\mu^{int}_{-}(m,t,z))\bigg),
\label{mua}
\end{flalign}
\begin{flalign}
\mu^{int}_{+}(n,t,z)=C^{int}(n,t)\bigg(I_4-Z^{-1}\sum_{m=n}^{\infty} \left(C^{int}(m,t)\right)^{-1}\hat{Z}^{n-m}(Q^{int}(m,t)\mu^{int}_{+}(m,t,z))\bigg).
\label{mub}
\end{flalign}
\label{mu}
\end{subequations}
These two solutions are related:
\begin{eqnarray}
\mu^{int}_{+}(n,t,z)=\mu^{int}_{-}(n,t,z)\hat{Z}^ne^{i\omega(z)t\hat{\Sigma}_3}s^{int}(z), ~~|z|=1, \label{sint}
\end{eqnarray}
where
\begin{eqnarray}
s^{int}(z)=\left( \begin{array}{cc} \tilde{a}^{int}(z) & b^{int}(z) \\ \tilde{b}^{int}(z) & a^{int}(z) \\ \end{array} \right),
\label{smint}
\end{eqnarray}
with $a^{int}(z)$, $b^{int}(z)$, $\tilde{a}^{int}(z)$, $\tilde{b}^{int}(z)$ being $2 \times 2$ diagonal-matrices.
Evaluating equation (\ref{sint}) as $n\rightarrow - \infty$ and at $t=0$, we find
\begin{eqnarray}
s^{int}(z)=\lim_{n\rightarrow -\infty}\hat{Z}^{-n}\mu^{int}_{+}(n,0,z).
\label{sfint}
\end{eqnarray}

\begin{theorem}\label{theorem2}
Given the initial data $\{q_0(n), p_0(n)\}$ belong to $l^1(\mathbb{Z})$ (here $\mathbb{Z}$ denotes the set of integers), define the associated spectral functions
$\{a^{int}(z),b^{int}(z),\tilde{a}^{int}(z),\tilde{b}^{int}(z)\}$  according to (\ref{smint}) and (\ref{sfint}).
Let $M^{int}(n, t, z)$ be the unique solution of the following RH problem\footnote{Again, we do not consider the possibility of the poles of $M(n, t, z)$ here.}
\begin{itemize}
\item $M^{int}(n, t, z)$ is an analytic function for $|z|>1$ and $|z|<1$.
\item \begin{eqnarray}
M^{int}_{-}(n,t,z)=M^{int}_{+}(n,t,z)J^{int}(n,t,z), \quad |z|=1,
\label{RHPint}
\end{eqnarray}
where $M^{int}(n,t,z)=M^{int}_{-}(n,t,z)$ for $|z|\leq 1$,
$M^{int}(n,t,z)=M^{int}_{+}(n,t,z)$ for $|z|\geq 1$, and
$J^{int}(n,t,z)$ is defined via the spectral functions
$\{a^{int}(z),b^{int}(z),\tilde{a}^{int}(z),\tilde{b}^{int}(z)\}$ by
\begin{eqnarray}
\begin{split}
J^{int}(n,t,z)&=\hat{Z}^ne^{i\omega(z)t\hat{\Sigma}_3}\left( \begin{array}{cc} I_{2} & -\gamma^{int}(z) \\
 \tilde{\gamma}^{int}(z) & I_{2}-\gamma^{int}(z)\tilde{\gamma}^{int}(z) \\ \end{array} \right),
\end{split}
\label{JMint}
\end{eqnarray}
with
\begin{eqnarray}
\gamma^{int}(z)=b^{int}(z)\left(\tilde{a}^{int}(z)\right)^{-1}, ~~\tilde{\gamma}^{int}(z)=\tilde{b}^{int}(z)\left(a^{int}(z)\right)^{-1}.
\label{gammaint}
\end{eqnarray}
\item \begin{eqnarray}
\begin{split}
&M^{int}(n,t,z)=I_{4}+\left( \begin{array}{cc} O_{2\times 2}(z^{-2},\text{even}) & O_{2\times 2}(z,\text{odd})  \\
  O_{2\times 2}(z^{-1},\text{odd}) & O_{2\times 2}(z^2,\text{even}) \\ \end{array} \right), \quad z\rightarrow (\infty,0).
\end{split}
\label{Maspint}
\end{eqnarray}
\end{itemize}
Define $Q^{int}(n,t)$, $P^{int}(n,t)$ by
\begin{eqnarray}
Q^{int}(n,t)=\lim_{z\rightarrow 0}(z^{-1}M^{int}(n+1,t,z))^{12}, \quad P^{int}(n,t)=\lim_{z\rightarrow \infty}(zM^{int}(n+1,t,z))^{21}.
\label{solutionint}
\end{eqnarray}
Then $Q^{int}(n,t)$, $P^{int}(n,t)$ solves the AL system (\ref{ALint}) with the initial condition
\begin{eqnarray}
\begin{split}
&Q^{int}(n,0)=\left( \begin{array}{cc} q_0(n) & 0 \\
 0 &  q_0(-n-1)\\ \end{array} \right),
\\
&P^{int}(n,0)=\left( \begin{array}{cc} p_0(n) & 0 \\
 0 &  p_0(-n-1)\\ \end{array} \right).
\end{split}
\label{ivint}
\end{eqnarray}
\end{theorem}

\subsubsection{ISM as a special case of UTM}
Equipped with the above results,
we are now able to demonstrate that the problem on the set of integers is a special case of
our $N=2$ graph problem.
We choose the initial data of the $N=2$ graph problem as
\begin{eqnarray}
\begin{split}
&q^1_0(n)=q_0(n),~~q^2_0(n)=q_0(-n-1),~~n\geq 0,
\\
&p^1_0(n)=p_0(n),~~p^2_0(n)=p_0(-n-1),~~n\geq 0,
\end{split}
\label{ivint}
\end{eqnarray}
and assume the two edges of the graph are connected via the boundary condition in the form of (\ref{bvint}).

\begin{lemma}\label{lemma2}
In the case of initial-boundary conditions satisfying (\ref{bvint}) and (\ref{ivint}),
the associated spectral matrices $s(z)$, $S(z)$ and $s^{int}(z)$ satisfy the following relations
\begin{subequations}
\begin{eqnarray}
&&s(z)=\Sigma s(\frac{1}{z}) \Sigma s^{int}(z),
\label{sredline}
\\
&&S(z)=\Sigma S(\frac{1}{z}) \Sigma,
\label{Sredline}
\end{eqnarray}
\label{sSredline}
\end{subequations}
where
\begin{eqnarray}
\Sigma=\sigma_3\otimes \sigma,
~~\sigma_3=\left( \begin{array}{cc} 1 & 0 \\
 0 &  -1 \\ \end{array} \right),
~~\sigma=\left( \begin{array}{cc} 0 & 1 \\
 1 &  0\\ \end{array} \right).
\label{Sigma}
\end{eqnarray}
\end{lemma}
{\bf Proof} \quad
Using symmetry relations
\begin{eqnarray}
\begin{split}
&U^{int}(n-1,t)=-\Sigma U^{int}(-n,t) \Sigma,
\\
&F^{int}(n-1,t)=\Sigma F^{int}(-n,t) \Sigma,
\end{split}
\label{UFsymmetry}
\end{eqnarray}
we can deduce that, if $\mu^{int}(n,t,z)$ solves (\ref{LPSint}), then so does $\Sigma \mu^{int}(-n,t,\frac{1}{z})\Sigma$.
This aspect and the uniqueness of normalized solutions implies that
\begin{eqnarray}
\mu_{-}^{int}(n,t,z)=\Sigma \mu_{+}^{int}(-n,t,\frac{1}{z}) \Sigma.
\label{muintsymmetry}
\end{eqnarray}
Using (\ref{muintsymmetry}) in (\ref{sint}), we find
\begin{eqnarray}
\mu^{int}_{+}(n,t,z)=\Sigma \mu^{int}_{+}(-n,t,\frac{1}{z})\Sigma \hat{Z}^ne^{i\omega(z)t\hat{\Sigma}_3}s^{int}(z), ~~|z|=1.
\label{muintsyms}
\end{eqnarray}
For $n\geq 0$, using (\ref{ivint}) we find that $\mu_{2}(n,0,z)$ and $\mu^{int}_{+}(n,0,z)$ satisfy the
same equation and have the same normalization as $n\rightarrow \infty$.
Thus, for $n\geq 0$, we have
\begin{eqnarray}
\mu_{2}(n,0,z)=\mu^{int}_{+}(n,0,z). \label{mu2int}
\end{eqnarray}
Evaluating equation (\ref{mu2int}) at $n=0$ and using $\mu_{2}(0,0,z)=s(z)$, we obtain
\begin{eqnarray}
s(z)=\mu^{int}_{+}(0,0,z). \label{smuint}
\end{eqnarray}
Evaluating equation (\ref{muintsyms}) at $n=t=0$ and using (\ref{smuint}), we find (\ref{sredline}).
We now turn to the proof of the symmetry (\ref{Sredline}).
Using (\ref{bvint}) we find that $\mu_{1}(0,t,z)$ and $\Sigma \mu_{1}(0,t,\frac{1}{z})\Sigma$ satisfy the
same equation and have the same normalization as $t\rightarrow 0$.
Thus, we have
\begin{eqnarray}
\mu_{1}(0,t,z)=\Sigma \mu_{1}(0,t,\frac{1}{z})\Sigma. \label{mu1sym}
\end{eqnarray}
Similarly, we find
\begin{eqnarray}
\mu_{3}(0,t,z)=\Sigma \mu_{3}(0,t,\frac{1}{z})\Sigma. \label{mu3sym}
\end{eqnarray}
Evaluating at $n=0$ for the following formula
\begin{eqnarray}
\mu_3(n,t,z)=\mu_1(n,t,z)\hat{Z}^ne^{i\omega(z)t\hat{\Sigma}_3}S(z),
\end{eqnarray}
we obtain
\begin{eqnarray}
\mu_3(0,t,z)=\mu_1(0,t,z)e^{i\omega(z)t\hat{\Sigma}_3}S(z).
\label{mu31n0}
\end{eqnarray}
Letting $z\mapsto \frac{1}{z}$ in (\ref{mu31n0}) and using the symmetries (\ref{mu1sym}) and (\ref{mu3sym}), we obtain
\begin{eqnarray}
\mu_3(0,t,z)=\mu_1(0,t,z)e^{i\omega(z)t\hat{\Sigma}_3}\left(\Sigma S(\frac{1}{z})\Sigma\right).
\label{mu31n0sym}
\end{eqnarray}
Equations (\ref{mu31n0}) and (\ref{mu31n0sym}) yield (\ref{Sredline}).
\QEDB

\begin{theorem}\label{theorem3}
Consider the RH problem defined in Theorem \ref{theorem1}
subject to $N=2$ and particular initial-boundary values satisfying (\ref{bvint}) and (\ref{ivint}).
Denote by $M^{red}(n, t, z)$ the unique solution of this RH problem and
let $Q^{red}(n, t)$, $P^{red}(n, t)$ be defined by (\ref{solution}) but with $M(n, t, z)$ replaced by $M^{red}(n, t, z)$.
Then
\begin{eqnarray}
Q^{int}(n,t)= Q^{red}(n,t), ~~P^{int}(n,t)= P^{red}(n,t),~~ n\in \mathbb{N}_0.
\label{Qredline}
\end{eqnarray}
\end{theorem}
{\bf Proof}~~
Define
\begin{eqnarray}
\begin{split}
\widetilde{M}^{red}(n,t,z)=\left\{\begin{array}{l} M^{red}(n,t,z), \quad z\in \bar{D}_{-in}\cup \bar{D}_{+out},
\\
M^{red}(n,t,z)J_1(n,t,z), \quad z\in \bar{D}_{+in},
\\
M^{red}(n,t,z)J^{-1}_3(n,t,z), \quad z\in \bar{D}_{-out}.
\end{array}\right.
\end{split}
\label{Mredtilde}
\end{eqnarray}
We find that $\widetilde{M}^{red}(n,t,z)$ only has a jump across the unit circle:
\begin{eqnarray}
\widetilde{M}_{-}^{red}(n,t,z)= \widetilde{M}_{+}^{red}J_2(n,t,z), ~~ |z|=1,
\label{MredtildeJ}
\end{eqnarray}
where
\begin{eqnarray}
J_2(n,t,z)=\hat{Z}^ne^{i\omega(z)t\hat{\Sigma}_3}\left( \begin{array}{cc} I_{2} & \tilde{\Gamma}(z)-\gamma(z) \\
 \tilde{\gamma}(z)-\Gamma(z) & I_{2}-(\tilde{\gamma}(z)-\Gamma(z))(\gamma(z)-\tilde{\Gamma}(z)) \\ \end{array} \right).
\end{eqnarray}
We can deduce that
\begin{subequations}
\begin{eqnarray}
 \tilde{\gamma}(z)-\Gamma(z)=\tilde{\gamma}^{int}(z),
 \label{gamaequa}
 \\
 \gamma(z)-\tilde{\Gamma}(z)=\gamma^{int}(z).
 \label{gamaequb}
\end{eqnarray}
\label{gamaequ}
\end{subequations}
Indeed, the symmetry (\ref{Sredline}) implies
$
\tilde{B}(z)\tilde{A}^{-1}(z)=-\sigma \tilde{B}(\frac{1}{z})\tilde{A}^{-1}(\frac{1}{z}) \sigma.
 \label{tildeR}
$
Using the global relation (\ref{gr5}), we obtain
$
 \tilde{B}(z)\tilde{A}^{-1}(z)=-\sigma \tilde{b}(\frac{1}{z})\tilde{a}^{-1}(\frac{1}{z}) \sigma.
 \label{tildeR1}
$
Using this relation we can write
\begin{eqnarray}
 \Gamma(z)=-a^{-1}(z)\sigma \tilde{b}(\frac{1}{z})\sigma \left(a(z)\sigma\tilde{a}(\frac{1}{z}) \sigma+b(z)\sigma\tilde{b}(\frac{1}{z}) \sigma\right)^{-1}.
 \label{Gamred}
\end{eqnarray}
Using (\ref{Gamred}) and the identity $a(z)\tilde{a}(z)-b(z)\tilde{b}(z)=I_2$, we can write
\begin{eqnarray}
\tilde{\gamma}(z)- \Gamma(z)=\left(\tilde{b}(z)\sigma\tilde{a}(\frac{1}{z}) \sigma+\tilde{a}(z)\sigma\tilde{b}(\frac{1}{z}) \sigma\right)
 \left(a(z)\sigma\tilde{a}(\frac{1}{z}) \sigma+b(z)\sigma\tilde{b}(\frac{1}{z}) \sigma\right)^{-1}.
 \label{gGamred}
\end{eqnarray}
The symmetry (\ref{sredline}) implies
\begin{eqnarray}
\begin{split}
\tilde{b}(z)\sigma\tilde{a}(\frac{1}{z}) \sigma+\tilde{a}(z)\sigma\tilde{b}(\frac{1}{z}) \sigma=\tilde{b}^{int}(z),
\\
 a(z)\sigma\tilde{a}(\frac{1}{z}) \sigma+b(z)\sigma\tilde{b}(\frac{1}{z}) \sigma=a^{int}(z).
\end{split}
\label{gGamred12}
\end{eqnarray}
Inserting (\ref{gGamred12}) into (\ref{gGamred}), we find (\ref{gamaequa}). Similar calculations yield (\ref{gamaequb}).
It follows from (\ref{gamaequ}) that
\begin{eqnarray}
J_2(n,t,z)=J^{int}(n,t,z).
\end{eqnarray}
Moreover, (\ref{Mredtilde}) implies that the normalization (\ref{Maspint}) also holds for $\widetilde{M}^{red}(n,t,z)$.
Therefore, $\widetilde{M}^{red}(n,t,z)$ and $M^{int}(n,t,z)$ satisfy exactly the same RH problem. Thus
\begin{eqnarray}
\widetilde{M}^{red}(n,t,z)= M^{int}(n,t,z), ~~ n\in \mathbb{N}_0.
\label{Mredline}
\end{eqnarray}
Finally, using
\begin{eqnarray}
\begin{split}
&\lim_{z\rightarrow 0}(z^{-1}\widetilde{M}^{red}(n,t,z))^{12}=\lim_{z\rightarrow 0}(z^{-1}M^{red}(n,t,z))^{12},
\\
&\lim_{z\rightarrow \infty}(z\widetilde{M}^{red}(n,t,z))^{21}=\lim_{z\rightarrow \infty}(zM^{red}(n,t,z))^{21},
\end{split}
\label{MtiM}
\end{eqnarray}
we find (\ref{Qredline}).
\QEDB

The formula (\ref{Qredline}) implies that the problem on the set of integers can be recovered from our $N=2$ matrix problem
on the set of non-negative integers as a special case by the following very intuitive formula
\begin{subequations}
\begin{eqnarray}
q(n,t)=\theta(n)q^1(n,t)+\theta(-n-1)q^2(-n-1,t), ~~n\in \mathbb{Z},
\label{qpa}
\\
p(n,t)=\theta(n)p^1(n,t)+\theta(-n-1)p^2(-n-1,t), ~~n\in \mathbb{Z},
\label{qpb}
\end{eqnarray}
\label{qp}
\end{subequations}
where $\theta(n)$ is defined by the following Heaviside theta function
\begin{eqnarray*}
\theta(n)=\left\{\begin{array}{l} 1, ~~n\geq 0,
\\
0,~~n<0.
\end{array}\right.
\end{eqnarray*}

\noindent
{\bf Remark 2.}
Our result presents a relationship between the ISM and UTM for the integrable DDEs.
Other aspects regarding the connection between the two approaches have been studied in \cite{BFS,BH3,VC2}.

\subsection {The IDNLS and nonlocal IDNLS equations as standard local reductions}
In the case of continuum problem, Caudrelier has established in \cite{VC2} that
both the classical NLS equation and the {\it nonlocal} NLS equation \cite{AM1,AMnon,AM2} can be obtained as standard {\it local} reductions
of a matrix AKNS system with an appropriate boundary condition.
Here we show this is also the case for the discrete problem:
both the IDNLS equation \cite{AL3} and the {\it nonlocal} IDNLS equation \cite{AMD}
can be obtained as standard {\it local} reductions of our matrix AL system on the non-negative integers
with the boundary values satisfying (\ref{bvint}).
Thus the ISM for the IDNLS equation and especially for the nonlocal IDNLS equation fit naturally within our results on the graph problem.

\subsubsection{Reductions in the potentials}
It can be checked directly that the matrix AL system (\ref{ALM}) in the case $N=2$ admits the following two local reductions between the potentials:
\begin{subequations}
\begin{eqnarray}
P(n,t)=\nu Q^{\dag}(n,t), ~~\nu=\pm 1,~~n\in \mathbb{N}_0,
\label{reduction1}
\\
P(n,t)=\nu\sigma Q^{\dag}(n,t)\sigma, ~~\nu=\pm 1, ~~n\in \mathbb{N}_0,
\label{reduction2}
\end{eqnarray}
\label{reduction}
\end{subequations}
where $Q^{\dag}(n,t)$ denotes the Hermitian of $Q(n,t)$, and $\sigma$ are defined by (\ref{sigma}).
We note that, as in the continuum case \cite{VC2},
the reductions (\ref{reduction}) admit an interpretation in the reduction group theory \cite{AVM}:
they arise as two different representations of a local $\mathbb{Z}_2$ reduction imposing on the $4\times 4$ Lax pair of the matrix AL equation (\ref{ALM}) in the case of $N=2$.

We first consider the reduction (\ref{reduction1}).
In this case, the matrix AL system (\ref{ALM}) in the case $N=2$ becomes
\begin{equation}
i\frac{dQ_{n}}{dt}+Q_{n+1}-2Q_{n}+Q_{n-1}-\nu Q^{\dag}_{n}Q_{n}\left( Q_{n+1}+ Q_{n-1}\right)=0.
 \label{ALMR1}
\end{equation}
The above equation holds for $n\in \mathbb{N}_0$ and $t>0$,
since we deal with an IBV problem for the lattice equation on the non-negative integers.
Note that we require that the reduction is compatible with the boundary condition (\ref{bvint}),
thus we can apply (\ref{qpa}), namely $Q_{n}=diag(q_{n},q_{-n-1})$, $n\in \mathbb{N}_0$, to (\ref{ALMR1}).
This in turn yields
\begin{subequations}
\begin{eqnarray}
i\frac{d q_n}{dt}+q_{n+1}-2q_{n}+q_{n-1}
-\nu |q_n|^2\left(q_{n+1}+q_{n-1}\right)=0, ~~n\in \mathbb{N}_0,
\label{idnlsa}
\\
i\frac{d q_{-n-1}}{dt}+ q_{-n-2} -2q_{-n-1}+q_{-n}-\nu |q_{-n-1}|^2\left(q_{-n-2}+q_{-n}\right)=0, ~n\in \mathbb{N}_0.
\label{idnlsb}
\end{eqnarray}
\label{idnlsab}
\end{subequations}
Equations (\ref{idnlsa}) and (\ref{idnlsb}) can be combined into
\begin{equation}
i\frac{d q_n}{dt}+q_{n+1}-2q_{n}+q_{n-1}-\nu |q_n|^2\left(q_{n+1}+q_{n-1}\right)=0, ~~n\in \mathbb{Z},
 \label{idnls}
\end{equation}
which is nothing but the IDNLS equation \cite{AL3} on the integers.
Next we consider the reduction (\ref{reduction2}).
In this case, the matrix AL system (\ref{ALM}) in the case $N=2$ becomes
\begin{equation}
i\frac{dQ_{n}}{dt}+Q_{n+1}-2Q_{n}+Q_{n-1}-\nu \sigma Q^{\dag}_n\sigma Q_n\left( Q_{n+1}+ Q_{n-1}\right)=0,
 \label{ALMR2}
\end{equation}
which holds for $n\in \mathbb{N}_0$ and $t>0$ as before.
By applying (\ref{qpa}), namely $Q_{n}=diag(q_{n},q_{-n-1})$, $n\in \mathbb{N}_0$, to (\ref{ALMR2}),
we obtain
\begin{subequations}
\begin{eqnarray}
i\frac{d q_n}{dt}+q_{n+1}-2q_{n}+q_{n-1}
-\nu q_{n}q^*_{-n-1}\left(q_{n+1}+q_{n-1}\right)=0, ~~n\in \mathbb{N}_0,
\label{nidnlsa}
\\
i\frac{d q_{-n-1}}{dt}+ q_{-n-2} -2q_{-n-1}+q_{-n}
-\nu q_{-n-1}q^*_{n}\left(q_{-n-2}+q_{-n}\right)=0, ~~n\in \mathbb{N}_0.
\label{nidnlsb}
 \end{eqnarray}
 \label{nidnlsab}
\end{subequations}
Equations (\ref{nidnlsa}) and (\ref{nidnlsb}) can be combined into the following nonlocal IDNLS equation on the integers:
\begin{equation}
i\frac{d q_n}{dt}+q_{n+1}-2q_{n}+q_{n-1}
-\nu q_{n}q^*_{-n-1}\left(q_{n+1}+q_{n-1}\right)=0, ~~n\in \mathbb{Z}.
 \label{nidnls}
\end{equation}

In summary, we find
\begin{proposition}\label{proposition3}
The IDNLS equation (\ref{idnls}) and the nonlocal IDNLS equation (\ref{nidnls}) can be obtained respectively as
standard {\it local} reductions (\ref{reduction1}) and (\ref{reduction2}) of the matrix AL system (\ref{ALM})
in the case of $N=2$ and the boundary condition satisfying (\ref{bvint}).
\end{proposition}

\noindent
{\bf Remark 3.} The nonlocal IDNLS equation (\ref{nidnls}) is slightly different from the following nonlocal IDNLS equation presented in \cite{AMD}:
\begin{equation}
i\frac{d q_n}{dt}+q_{n+1}-2q_{n}+q_{n-1}
-\nu q_{n}q^*_{-n}\left(q_{n+1}+q_{n-1}\right)=0.
 \label{nidnls2}
\end{equation}
The nonlocal term appearing in the nonlinear terms of our equation (\ref{nidnls}) is $q^*(-n-1,t)$,
while the nonlocal term appearing in the nonlinear terms of equation (\ref{nidnls2}) is $q^*(-n,t)$.
We note that, in general, the nonlocal term can be taken as $q^*(-n-n_0,t)$, and the resulting nonlocal IDNLS equation reads
\begin{equation}
i\frac{d q_n}{dt}+q_{n+1}-2q_{n}+q_{n-1}
-\nu q_{n}q^*_{-n-n_0}\left(q_{n+1}+q_{n-1}\right)=0, 
 \label{gnidnls}
\end{equation}
where $n_0$ is an arbitrary fixed integer.
Indeed, the AL system (\ref{aleq1}) admits a general nonlocal reduction $p(n,t)=\nu q^*(-n-n_0,t)$, $\nu=\pm 1$.
Using this reduction, we immediately find the above general nonlocal IDNLS equation (\ref{gnidnls}).

\subsubsection{Reduction symmetries in the spectral functions}

The reductions in the potentials induce important symmetries in the spectral functions.
For the nonlocal IDNLS equation, it was shown in \cite{AMD} that the resulting symmetries of the spectral functions are very different from
those of classical IDNLS equation.
We will show in the following that these two different symmetries associated with the two equations also appear naturally
from the reductions (\ref{reduction}) imposed on the matrix AL system (\ref{ALM}) in the case $N=2$.

We first write the symmetry reductions (\ref{reduction}) as
\begin{eqnarray}
U(n,t)=-D U^{\dag}(n,t)D,
\label{Ureduction}
\end{eqnarray}
where, for the IDNLS equation (\ref{idnls}),
\begin{eqnarray}
D=\left( \begin{array}{cc} -\nu I_2 & 0 \\
0 &  I_2 \\ \end{array} \right),
\label{DIDNLS}
\end{eqnarray}
while for the nonlocal IDNLS equation (\ref{nidnls}),
\begin{eqnarray}
D=\left( \begin{array}{cc} -\nu \sigma & 0 \\
0 &  \sigma \\ \end{array} \right).
\label{DNIDNLS}
\end{eqnarray}

\begin{lemma}\label{lemma3}
Under the reduction (\ref{Ureduction}),
the associated spectral matrices $s(z)$ and $S(z)$ satisfy symmetry relations:
\begin{subequations}
\begin{eqnarray}
&&s^{-1}(z)=D s^{\dag}(\frac{1}{z^*}) D,
\label{ssinv}
\\
&&S^{-1}(z)=D S^{\dag}(\frac{1}{z^*}) D.
\label{SSinv}
\end{eqnarray}
\label{sSinv}
\end{subequations}
\end{lemma}
{\bf Proof}
It can be checked directly that under the reduction (\ref{Ureduction}), if $\mu(n,t,z)$ solves (\ref{LPSTM}) in the case $N=2$,
then $\mu^{-1}(n,t,z)$ and $D\mu^{\dag}(n,t,\frac{1}{z^*})D$ satisfy the same equations
\begin{subequations}
\begin{eqnarray}
&&\Phi(n+1,t,z)F(n,t)-\hat{Z}\Phi(n,t,z)=-Z\Phi(n,t,z)U(n,t),
\label{LPSMinv}\\
&&\Phi_t(n,t,z)-i\omega(z)[\Sigma_3,\Phi(n,t,z)]=-\Phi(n,t,z)V(n,t,z).
\label{LPTMinv}
\end{eqnarray}
\label{LPSTMinv}
\end{subequations}
Applying this to $\mu_2(n, 0, z)$ and using uniqueness of a normalized solution, we find
\begin{eqnarray}
\mu^{-1}_2(n, 0, z)=D\mu^{\dag}_2(n,0,\frac{1}{z^*})D.
\label{mu2inv}
\end{eqnarray}
Applying this to $\mu_3(0, t, z)$ and using uniqueness of a normalized solution, we find
\begin{eqnarray}
\mu^{-1}_3(0, t, z)=D\mu^{\dag}_3(0,t,\frac{1}{z^*})D.
\label{mu3inv}
\end{eqnarray}
Evaluating equation (\ref{mu2inv}) at $n=0$, we find (\ref{ssinv}).
Evaluating equation (\ref{mu3inv}) at $t=0$, we find (\ref{SSinv}).
\QEDB

\begin{proposition}\label{proposition4}
The matrix of spectral functions of the IDNLS equation (\ref{idnls}) is of the generic form
\begin{eqnarray}
s_{IDNLS}(z)=\left( \begin{array}{cc} \left(\alpha(\frac{1}{z^*})\right)^{*}  & \beta(z)
\\ \nu \left(\beta(\frac{1}{z^*})\right)^{*} & \alpha(z) \\ \end{array} \right).
\label{smintIDNLS}
\end{eqnarray}
The matrix of spectral functions of the nonlocal IDNLS equation (\ref{nidnls}) is of the generic form
\begin{eqnarray}
s_{NIDNLS}(z)=\left( \begin{array}{cc} \tilde{\alpha}(z) & \beta(z)
\\ \nu \left(\beta(z^*)\right)^{*} & \alpha(z) \\ \end{array} \right),
\label{smintNIDNLS}
\end{eqnarray}
where $\alpha(z)$ and $\tilde{\alpha}(z)$ satisfy extra symmetries
\begin{eqnarray}
\begin{split}
&\alpha(z)=\left(\alpha(z^*)\right)^*,
\\
&\tilde\alpha(z)=\left(\tilde\alpha(z^*)\right)^*.
\end{split}
\label{symNIDNLS}
\end{eqnarray}
\end{proposition}
{\bf Proof}
From (\ref{sredline}), we find
\begin{eqnarray}
\left(s^{int}(z)\right)^{-1}=\Sigma s^{int}(\frac{1}{z}) \Sigma.
\label{sredlineinv}
\end{eqnarray}
The formula (\ref{ssinv}) implies
\begin{eqnarray}
\left(s^{int}(z)\right)^{-1}=D \left(s^{int}(\frac{1}{z^*})\right)^{\dag} D.
\label{slineinv}
\end{eqnarray}
For $D$ being defined by (\ref{DIDNLS}), from (\ref{sredlineinv}) and (\ref{slineinv}) we find
\begin{eqnarray}
s^{int}(z)=\left(
\begin{array}{cc|cc}
	\left(a^{int}(\frac{1}{z^*})\right)^{*} & 0 & b^{int}(z) & 0 \\
	0 & a^{int}(\frac{1}{z}) & 0 & b^{int}(\frac{1}{z}) \\ \hline
    \nu \left(b^{int}(\frac{1}{z^*})\right)^{*} & 0 & a^{int}(z) & 0 \\
    0 & \nu \left(b^{int}(z^*)\right)^{*} & 0 & \left(a^{int}(z^*)\right)^{*}
\end{array}
\right).
\end{eqnarray}
In view of the structure of (\ref{fQPint}), from the above redundant $4\times 4$ matrix we extract the following $2\times 2$
matrix for spectral functions of the IDNLS equation (\ref{idnls}):
\begin{eqnarray}
s_{IDNLS}(z)=\left( \begin{array}{cc} \left(a^{int}(\frac{1}{z^*})\right)^{*}  & b^{int}(z)
\\ \nu \left(b^{int}(\frac{1}{z^*})\right)^{*} & a^{int}(z) \\ \end{array} \right).
\label{smintIDNLS1}
\end{eqnarray}
For $D$ being defined by (\ref{DNIDNLS}), from (\ref{sredlineinv}) and (\ref{slineinv}) we find
\begin{eqnarray}
s^{int}(z)=\left(
\begin{array}{cc|cc}
	\tilde{a}^{int}(z) & 0 & b^{int}(z) & 0 \\
	0 & a^{int}(\frac{1}{z}) & 0 & b^{int}(\frac{1}{z}) \\ \hline
    \nu \left(b^{int}(z^*)\right)^{*} & 0 & a^{int}(z) & 0 \\
    0 & \nu \left(b^{int}(\frac{1}{z^*})\right)^{*} & 0 & \tilde{a}^{int}(\frac{1}{z})
\end{array}
\right),
\end{eqnarray}
with
\begin{eqnarray}
\begin{split}
&a^{int}(z)=\left(a^{int}(z^*)\right)^*,
\\
&\tilde{a}^{int}(z)=\left(\tilde{a}^{int}(z^*)\right)^*.
\end{split}
\label{symNIDNLS1}
\end{eqnarray}
In view of the structure of (\ref{fQPint}), from the above redundant $4\times 4$ matrix we extract the following $2\times 2$
matrix for spectral functions of the nonlocal IDNLS equation (\ref{nidnls}):
\begin{eqnarray}
s_{NIDNLS}(z)=\left( \begin{array}{cc} \tilde{a}^{int}(z) & b^{int}(z)
\\ \nu \left(b^{int}(z^*)\right)^{*} & a^{int}(z) \\ \end{array} \right),
\label{smintNIDNLS1}
\end{eqnarray}
where $a^{int}(z)$ and $\tilde{a}^{int}(z)$ satisfy extra symmetries (\ref{symNIDNLS}).
\QEDB

\subsection {AL lattice system with an integrable defect}

In classical integrable field theories, a defect for an integrable system in space can be viewed
as an internal boundary condition on the fields and their time and space
derivatives at a given point (defect location) \cite{BCZ1,BCZ2,CZ1,CZ2}.
A crucial observation was that the integrability of a defect system could survive
if one defines defect boundary conditions for the system as B\"{a}cklund transformations (BTs)
frozen at the defect location \cite{CZ2}.
Using this observation, a generating function for the defect contributions to the
conserved quantities was explicitly constructed in \cite{VC3} for integrable PDEs associated with AKNS spectral problem,
and furthermore the integrability of the defect system was studied in \cite{CK}.
However, for the integrable DDEs, to our knowledge, the analogous result for the AL lattice system (\ref{aleq1}) has not been reported in the literature.
We will present this important result in Appendix A.

Here we note that the problem on the line with a defect fits within the framework of the problem on a simple graph.
Indeed, for the problem in the continuum case, the line with a defect at a fixed point can be seen as a 
graph with two half-lines attached to a vertex,
and the associated defect condition can be interpreted as a boundary condition representing a connection between two edges of the graph \cite{VC1}.
For the problem in the discrete case, by freezing at $n=0$ the BT (\ref{BT}) of the AL system derived in Appendix A,
we find the following defect boundary condition
\begin{eqnarray}
\begin{split}
&c_1\left(E(t)\right)^2g_0^1(t)-c_2g_0^2(t)=c_4\left(E(t)\right)^2g_{-1}^2(t)-c_3g_{-1}^1(t),
\\
&c_1\left(E(t)\right)^2h_{-1}^2(t)-c_2h_{-1}^1(t)=c_4\left(E(t)\right)^2h_{0}^1(t)-c_3h_{0}^2(t),
\\
&c_1\left(E(t)\right)^2\dot{g}_0^1(t)-c_2\dot{g}_0^2(t)
\\=&i\bigg\{\left(c_2+c_4\left(E(t)\right)^2\right)\left(g_0^2(t)-g_{-1}^2(t)\right)-\left(c_3+c_1\left(E(t)\right)^2\right)\left(g_0^1(t)-g_{-1}^1(t)\right)
\\& -c_1\left(E(t)\right)^2g_0^1(t)\left(g_{-1}^1(t)h_0^1(t)+g_0^2(t)h_{-1}^2(t)\right)+c_2g_0^2(t)\left(g_{0}^1(t)h_{-1}^1(t)+h_0^2(t)g_{-1}^2(t)\right)
\bigg\},
\\
&c_4\left(E(t)\right)^2\dot{h}_{0}^1(t)-c_3\dot{h}_{0}^2(t)
\\=&i\bigg\{\left(c_3+c_1\left(E(t)\right)^2\right)\left(h_{-1}^2(t)-h_{0}^2(t)\right)+\left(c_2+c_4\left(E(t)\right)^2\right)\left(h_0^1(t)-h_{-1}^1(t)\right)
\\& +c_4\left(E(t)\right)^2h_0^1(t)\left(g_0^1(t)h_{-1}^1(t)+g_{-1}^2(t)h_0^2(t)\right)-c_3h_0^2(t)\left(h_{0}^1(t)g_{-1}^1(t)+g_0^2(t)h_{-1}^2(t)\right)
\bigg\},
\end{split}
\label{BVD0}
\end{eqnarray}
where the dot denotes derivative with respect to time, $c_1$, $c_2$, $c_3$ and $c_4$ are arbitrary constants, and
$$E(t)=\exp\left\{\frac{i}{2}\int_{0}^t\left( g_0^2(\tau)h_{-1}^2(\tau)-g_0^1(\tau)h_{-1}^1(\tau)-h_0^2(\tau)g_{-1}^2(\tau)+h_0^1(\tau)g_{-1}^1(\tau)\right)d\tau\right\}.$$
In our present context, this defect condition represents a connection
between two semi-infinite lattices of the $N=2$ graph.
In particular, for the IDNLS equation (\ref{idnls}), the additional symmetry $p(n,t)=\nu q^*(n,t)$ reduces the above defect condition to
\begin{eqnarray}
\begin{split}
&c_1\left(E(t)\right)^2g_0^1(t)-c_2g_0^2(t)=c_1^*\left(E(t)\right)^2g_{-1}^2(t)-c_2^*g_{-1}^1(t),
\\
&c_1\left(E(t)\right)^2\dot{g}_0^1(t)-c_2\dot{g}_0^2(t)
\\=&i\bigg\{\left(c_2+c_1^*\left(E(t)\right)^2\right)\left(g_0^2(t)-g_{-1}^2(t)\right)-\left(c_2^*+c_1\left(E(t)\right)^2\right)\left(g_0^1(t)-g_{-1}^1(t)\right)
\\& -c_1\left(E(t)\right)^2g_0^1(t)\left(g_{-1}^1(t)h_0^1(t)+g_0^2(t)h_{-1}^2(t)\right)+c_2g_0^2(t)\left(g_{0}^1(t)h_{-1}^1(t)+h_0^2(t)g_{-1}^2(t)\right)
\bigg\},
\end{split}
\label{BVD1}
\end{eqnarray}
where $c_1$ and $c_2$ are arbitrary constants, and
$$E(t)=\exp\left\{-\nu\int_{0}^t {\rm Im}\left(g_0^2(\tau)\left(g_{-1}^2(\tau)\right)^*-g_0^1(\tau)\left(g_{-1}^1(\tau)\right)^*\right)d\tau\right\}.$$
The defect conditions (\ref{BVD0}) and (\ref{BVD1}) look very complicated,
however they emerge naturally from the BT of AL system.
This implies that specific soliton solutions may be constructed explicitly.
For the problem in the continuum case, this was done for the NLS equation in \cite{CZ2} by direct ansatz on the one and two soliton solutions.
The analogous issue for the present discrete case will be studied in the future.

\subsection{Case $N \geq 2$: Kirchhoff boundary conditions}
The connections between the edges of the graph presented in previous subsections are in the case of $N=2$.
For $N \geq 2$, an interesting nontrivial connection between the $N$ edges of a graph is described by the so-called Kirchhoff boundary conditions \cite{ACFD1,KS,Nov}.
By analogy with the integrable PDEs case, in the simple case of a Kirchhoff boundary condition \cite{ACFD1},
the connection between the edges of the graph $\mathcal{G}$ for the AL system is given by
\begin{eqnarray}
\begin{split}
g^1_0(t)=g^2_{-1}(t)=g^3_{-1}(t)=\cdots=g^N_{-1}(t),~~h^1_0(t)=h^2_{-1}(t)=h^3_{-1}(t)=\cdots=h^N_{-1}(t),
\\
\sum_{j=1}^N \left(g^j_0(t)-g^j_{-1}(t)\right)=0,~~\sum_{j=1}^N \left(h^j_0(t)-h^j_{-1}(t)\right)=0.
\end{split}
\label{KB}
\end{eqnarray}
In particular, for $N=2$, (\ref{KB}) becomes
\begin{eqnarray}
\begin{split}
g^1_0(t)=g^2_{-1}(t),~~h^1_0(t)=h^2_{-1}(t),
\\
\sum_{j=1}^2 \left(g^j_0(t)-g^j_{-1}(t)\right)=0,~~\sum_{j=1}^2 \left(h^j_0(t)-h^j_{-1}(t)\right)=0,
\end{split}
\label{KB2}
\end{eqnarray}
which coincides with the boundary condition (\ref{bvint}).
For $N=3$, (\ref{KB}) becomes
\begin{eqnarray}
\begin{split}
g^1_0(t)=g^2_{-1}(t)=g^3_{-1}(t),~~h^1_0(t)=h^2_{-1}(t)=h^3_{-1}(t),
\\
\sum_{j=1}^3 \left(g^j_0(t)-g^j_{-1}(t)\right)=0,~~\sum_{j=1}^3 \left(h^j_0(t)-h^j_{-1}(t)\right)=0,
\end{split}
\label{KB3}
\end{eqnarray}
which defines a boundary condition for the AL system on a graph constituted of trilete lattices \cite{Discrete3,Discrete4}.
We note that, for the problem in continuum case, the analogous boundary problem 
on the $N=3$ star-graph (called $Y$-junction in the literature) is of special physical interest \cite{App2}.
For the NLS equation, the global well-posedness and the behaviour of solitary wave solution of this problem
were studied recently in \cite{ACFD1}.
It is unclear whether similar considerations can be applied to the present discrete case.


\section {Concluding remarks}

We have presented an approach to analyze IBV problems for integrable DDEs on a graph that is composed of $N$ semi-infinite lattices (edges).
As in the continuum case, we first formulated the problem on the graph into a certain matrix IBV problem.
Then we analyzed such a matrix IBV problem by extending the UTM for integrable DDEs in the scalar case to the one in the matrix case.
We also discussed the connections between the edges of the graph and compared our results with some previously known studies.
In particular, we discussed three interesting boundary conditions that represent nontrivial connections between the edges of the graph.
The first one, corresponding to a linearizable boundary condition for $N=2$, was given by (\ref{bvint}).
Regarding this boundary condition, we demonstrated in detail that how our results reproduce the standard ISM on the set of integers as particular case
and how a nonlocal reduction of an integrable DDE can be obtained as a standard {\it local} reduction.
The second one, corresponding to an integrable defect boundary condition, was given by (\ref{BVD0});
the third one, corresponding to a discrete analogue of Kirchhoff type boundary conditions for integrable PDEs, was given by (\ref{KB}).
Regarding these two boundary conditions, despite some impressive results on the problems in continuum case \cite{CZ2,VC3,ACFD1},
the study of the related problems in the present discrete case is just at its beginning
(we addressed several interesting issues concerning this subject at the end of Section 4.3 and Section 4.4).

Although the approach was illustrated for the AL lattice system,
it is clear that the same approach can be applied to other integrable nonlinear DDEs.


\section*{ACKNOWLEDGMENTS}

This work was supported by the National Natural Science Foundation of China (Grant No. 11771186).

\begin{appendices}

\section{AL lattice system with an integrable defect}

We now derive a defect condition for the AL system that can preserve the integrability of the system.
This result provides a discrete analogue of the integrable defect condition for the integrable PDEs \cite{VC3}.
We first derive a BT for AL system.
Then, by fixing the BT at $n=0$, we obtain the desired integrable defect boundary conditions.
Finally, we construct explicitly a generating function for the infinite number of conserved quantities for the defect AL system.

\subsection {B\"{a}cklund transformations for the AL system}

Let $q^1(n,t)$, $p^1(n,t)$ and $q^2(n,t)$, $p^2(n,t)$ be two solutions of the AL system (\ref{aleq1}) for $n\in \mathbb{Z}$;
let $\Phi^j(n, t, z)=\left(\Phi_1^j(n, t, z), \Phi_2^j(n, t, z)\right)^T$, $j=1,2$, be the corresponding eigenfunctions
that satisfy the following auxiliary problems, for $j=1,2$,
\begin{subequations}
\begin{eqnarray}
&&\Phi^j(n+1,t,z)=\mathcal{W}_j(n,t,z)\Phi^j(n,t,z),
\label{LPSBT}\\
&&\Phi^j_t(n,t,z)=\mathcal{T}_j(n,t,z)\Phi^j(n,t,z),
\label{LPTBT}
\end{eqnarray}
\label{LPSTBT}
\end{subequations}
where
\begin{subequations}
\begin{eqnarray}
&&\mathcal{W}_j(n,t,z)=\frac{1}{f^j(n,t)}(\mathcal{Z}+\mathcal{U}^{j}(n,t)),
\label{W}
\\
&&\mathcal{T}_j(n,t,z)=i\omega(z)\sigma_3+\mathcal{V}^j(n,t,z),
\label{T}
\end{eqnarray}
\label{WT}
\end{subequations}
with $\mathcal{Z}$, $f^j(n,t)$, $\mathcal{U}^{j}(n,t)$, $\mathcal{V}^j(n,t,z)$, $j=1,2$, being defined by (\ref{pqh}).
Suppose that $\Phi^1(n, t, z)$ and $\Phi^2(n, t, z)$ are related by gauge transformation
\begin{eqnarray}
\Phi^2(n,t,z)=\mathcal{D}(n, t, z)\Phi^1(n, t, z),
\label{GT}
\end{eqnarray}
where $D(n, t, z)$ is a $2\times 2$ matrix.
The matrix $D(n, t, z)$ satisfies the following equations
\begin{subequations}
\begin{eqnarray}
\mathcal{D}(n+1, t, z)\mathcal{W}^{1}(n,t,z)=\mathcal{W}^{2}(n,t,z)\mathcal{D}(n,t,z),
\label{DEa}
\\
\mathcal{D}_t(n,t,z)=\mathcal{T}^2(n,t,z)\mathcal{D}(n,t,z)-\mathcal{D}(n,t,z)\mathcal{T}^1(n,t,z).
\label{DEb}
\end{eqnarray}
\label{DE}
\end{subequations}
We look for $D(n, t, z)$ in the form of
\begin{eqnarray}
\mathcal{D}(n, t, z)=z\mathcal{D}_2(n, t)+\frac{1}{z}\mathcal{D}_1(n, t)+\mathcal{D}_0(n, t),
\label{DTDF}
\end{eqnarray}
where the $2 \times 2$ matrices $\mathcal{D}_0(n, t)$, $\mathcal{D}_1(n, t)$ and $\mathcal{D}_2(n, t)$ are dependent on $q^j(n,t)$, $p^j(n,t)$, $j=1,2$,
but independent on $z$.
By substituting (\ref{DTDF}) into (\ref{DE}) and equating the coefficients of powers of $z$, we find
\begin{lemma} \label{lemma4}
Let a solution of (\ref{DE}) be in the form of (\ref{DTDF}). Then
\begin{equation}
\mathcal{D}_2=\left( \begin{array}{cc} d^{11}_2(n,t) & 0
\\ 0 & d^{22}_2(n,t) \\ \end{array} \right),
~
\mathcal{D}_1=\left( \begin{array}{cc} d^{11}_1(n,t) & 0
\\ 0 & d^{22}_1(n,t) \\ \end{array} \right),
~
\mathcal{D}_0=\left( \begin{array}{cc} 0 & d^{12}_0(n,t)
\\ d^{21}_0(n,t) & 0 \\ \end{array} \right),
\label{DT2}
\end{equation}
where, for all $n\in \mathbb{Z}$,
\begin{eqnarray}
\begin{split}
&d^{11}_2(n,t)=c_1E(t)\Delta(n,t), ~~ d^{22}_2(n,t)=\frac{c_2}{E(t)\Delta(n,t)},
\\
&d^{11}_1(n,t)=\frac{c_3}{E(t)\Delta(n,t)}, ~~d^{22}_1(n,t)=c_4E(t)\Delta(n,t),
\\
&d^{12}_0(n,t)=d^{11}_2(n,t)q^1(n,t)-d^{22}_2(n,t)q^2(n,t)=d^{22}_1(n,t)q^2(n-1,t)-d^{11}_1(n,t)q^1(n-1,t),
\\
&d^{21}_0(n,t)=d^{11}_2(n,t)p^2(n-1,t)-d^{22}_2(n,t)p^1(n-1,t)=d^{22}_1(n,t)p^1(n,t)-d^{11}_1(n,t)p^2(n,t),
\end{split}
\label{d21}
\end{eqnarray}
with
$c_1$, $c_2$, $c_3$, $c_4$ being arbitrary constants, $\Delta(0,t)=1$, and
\begin{eqnarray}
\begin{split}
&\Delta(n,t)=\prod_{m=0}^{n-1}\frac{\left(1-q^1(m,t)p^1(m,t)\right)^{\frac{1}{2}}}{\left(1-q^2(m,t)p^2(m,t)\right)^{\frac{1}{2}}},
~\Delta(-n,t)=\prod_{m=-n}^{-1}\frac{\left(1-q^2(m,t)p^2(m,t)\right)^{\frac{1}{2}}}{\left(1-q^1(m,t)p^1(m,t)\right)^{\frac{1}{2}}},
~ n\geq 1,
\\
&E(t)=e^{\frac{i}{2}\int_{0}^t\left( q^2(0,\tau)p^2(-1,\tau)-q^1(0,\tau)p^1(-1,\tau)-p^2(0,\tau)q^2(-1,\tau)+p^1(0,\tau)q^1(-1,\tau)\right)d\tau}.
\end{split}
\label{Delta}
\end{eqnarray}
\end{lemma}

\begin{lemma} \label{lemma5}
(BT for the AL system)
Let $q^1(n,t)$, $p^1(n,t)$ and $q^2(n,t)$, $p^2(n,t)$ be two solutions of the AL system (\ref{aleq1});
let the corresponding eigenfunctions be related by gauge transformation (\ref{GT})
with $\mathcal{D}(n, t, z)$ being given by Lemma \ref{lemma4}.
Then $q^1(n,t)$, $p^1(n,t)$ and $q^2(n,t)$, $p^2(n,t)$ are related by the following difference and differential equations
\begin{eqnarray}
\begin{split}
&d^{11}_2(n,t)q^1(n,t)-d^{22}_2(n,t)q^2(n,t)=d^{22}_1(n,t)q^2(n-1,t)-d^{11}_1(n,t)q^1(n-1,t),
\\
&d^{22}_1(n,t)p^1(n,t)-d^{11}_1(n,t)p^2(n,t)=d^{11}_2(n,t)p^2(n-1,t)-d^{22}_2(n,t)p^1(n-1,t),
\\
&d^{11}_2(n,t)q^1_t(n,t)-d^{22}_2(n,t)q^2_t(n,t)
\\=&i\big\{\left(d^{22}_2(n,t)+d^{22}_1(n,t)\right)\left(q^2(n,t)-q^2(n-1,t)\right) -\left(d^{11}_2(n,t)+d^{11}_1(n,t)\right)\left(q^1(n,t)-q^1(n-1,t)\right)
\\& -d^{11}_2(n,t)q^1(n,t)\left(q^1(n-1,t)p^1(n,t)+q^2(n,t)p^2(n-1,t)\right)
\\&+d^{22}_2(n,t)q^2(n,t)\left(q^1(n,t)p^1(n-1,t)+p^2(n,t)q^2(n-1,t)\right)
\big\},
\\
&d^{22}_1(n,t)p_t^1(n,t)-d^{11}_1(n,t)p_t^2(n,t)
\\=&i\big\{\left(d^{11}_2(n,t)+d^{11}_1(n,t)\right)\left(p^2(n-1,t)-p^2(n,t)\right) +\left(d^{22}_2(n,t)+d^{22}_1(n,t)\right)\left(p^1(n,t)-p^1(n-1,t)\right)
\\& +d^{22}_1(n,t)p^1(n,t)\left(q^1(n,t)p^1(n-1,t)+p^2(n,t)q^2(n-1,t)\right)
\\&-d^{11}_1(n,t)p^2(n,t)\left(p^1(n,t)q^1(n-1,t)+q^2(n,t)p^2(n-1,t)\right)
\big\},
\end{split}
\label{BT}
\end{eqnarray}
where $d^{jj}_k(n,t)$, $j,k=1,2$ are given by (\ref{d21}).
\end{lemma}
{\bf Proof}
By substituting (\ref{DTDF}) and (\ref{DT2}) into (\ref{DE}), we find that the off-diagonal entries of the resulting equation of (\ref{DEa}) yield the first two equations of (\ref{BT}), while the off-diagonal entries of the resulting equation of (\ref{DEb}) yield the last two equations of (\ref{BT}).
\QEDB

\subsection{Defect conditions as frozen B\"{a}cklund transformations}

As in the continuum case,
we define defect conditions for the AL system (\ref{aleq1}) as the BT (\ref{BT}) ``frozen" at $n=n_0$, the position of the defect.
More precisely, for a fixed point $n_0\in \mathbb{Z}$
we suppose $q^1(n,t)$, $p^1(n,t)$ satisfies the AL lattice system (\ref{aleq1}) for $n<n_0$;
$q^2(n,t)$, $p^2(n,t)$ satisfies the AL system (\ref{aleq1}) for $n>n_0$;
at $n=n_0$, they are connected by the relations (\ref{BT}).
For the associated auxiliary problem,
we suppose the system (\ref{LPSTBT}) with $j=1$ exists for $n<n_0$, while the one with $j=2$ exists for $n>n_0$,
and the two systems are connected by the relations (\ref{GT}) at $n=n_0$.
Following the terminology of the analogous problem for integrable PDEs \cite{BCZ1,BCZ2,CZ1,CZ2,VC3,CK},
we refer to the matrix $\mathcal{D}(n_0, t, z)$ as the defect matrix and the relations (\ref{BT}) at $n=n_0$ as defect conditions.

We note that such defect conditions preserve the integrability of the AL system in the sense that there exists an infinite set of conserved quantities.
Indeed, we find the following conclusion.
\begin{proposition}\label{pro5}
Let $\Omega^j=\frac{\Phi_2^j(n, t, z)}{\Phi_1^j(n, t, z)}$, $j=1,2$, and let
\begin{equation}
I(z)=I^{left}_{bulk}(z)+I^{right}_{bulk}(z)+I_{defect}(z),
\label{GI}
\end{equation}
where
\begin{subequations}
\begin{eqnarray}
&&I^{left}_{bulk}(z)=\sum_{n=-\infty}^{n_0-1} \ln \left(\left(f^1(n,t)\right)^{-1}\left(1+z^{-1}q^1(n,t)\Omega^1\right)\right),
\label{IL}
\\
&&I^{right}_{bulk}(z)=\sum_{n=n_0}^{\infty} \ln \left(\left(f^2(n,t)\right)^{-1}\left(1+z^{-1}q^2(n,t)\Omega^2\right)\right),
\label{IR}
\\
&&I_{defect}(z)=\left.\ln\left(\mathcal{D}^{11}+\mathcal{D}^{12}\Omega^1\right)\right|_{n=n_0},
\label{ID}
\end{eqnarray}
\label{I}
\end{subequations}
and $\mathcal{D}^{ij}$, $i,j=1,2$, are the $ij$-entries of the defect matrix $\mathcal{D}(n,t,z)$.
Then
\begin{equation}
I_t(z)=0.
\label{It}
\end{equation}
\end{proposition}
{\bf Proof} ~~From (\ref{LPSTBT}), we find
\begin{subequations}
\begin{eqnarray}
\left(\left(\mathcal{S}-1\right)\ln \Phi^j_1(n,t,z)\right)_t=\left(\ln\left(\left(f^j(n,t)\right)^{-1}\left(z+q^j(n,t)\Omega^j\right)\right)\right)_t, ~~j=1,2,
\label{cln}
\\
\left(\mathcal{S}-1\right)\left(\ln \Phi^j_1(n,t,z)\right)_t=\left(\mathcal{S}-1\right)\left(\mathcal{T}^{11}_j+\mathcal{T}^{12}_j\Omega^j\right), ~~j=1,2,
\label{clt}
\end{eqnarray}
\label{clny}
\end{subequations}
where $\mathcal{S}$ denotes the shift operator, i.e. $\mathcal{S}f(n)=f(n+1)$,
and the quantities $\mathcal{T}^{kl}_j$, $k,l=1,2$, denote $kl$-entries of the  matrix $\mathcal{T}_j(n,t,z)$.
Then, $\left(\left(\mathcal{S}-1\right)\ln \Phi^j_1(n,t,z)\right)_t=\left(\mathcal{S}-1\right)\left(\ln \Phi^j_1(n,t,z)\right)_t$
yields
\begin{subequations}
\begin{eqnarray}
\left(\ln\left(\left(f^1(n,t)\right)^{-1}\left(z+q^1(n,t)\Omega^1\right)\right)\right)_t=\left(\mathcal{S}-1\right)\left(\mathcal{T}^{11}_1+\mathcal{T}^{12}_1\Omega^1\right),
~~n<n_0,
\label{clL}
\\
\left(\ln\left(\left(f^2(n,t)\right)^{-1}\left(z+q^2(n,t)\Omega^2\right)\right)\right)_t=\left(\mathcal{S}-1\right)\left(\mathcal{T}^{11}_2+\mathcal{T}^{12}_2\Omega^2\right),
~~n>n_0.
\label{clR}
\end{eqnarray}
\label{clLR}
\end{subequations}
Using the above relations and the rapid decay of $q^j(n,t)$ and $p^j(n,t)$, $j=1,2$, we obtain
\begin{eqnarray}
\begin{split}
\left(I^{left}_{bulk}(z)+I^{right}_{bulk}(z)\right)_t=\left.\left(\mathcal{T}^{11}_1+\mathcal{T}^{12}_1\Omega^1
-\mathcal{T}^{11}_2-\mathcal{T}^{12}_2\Omega^2\right)\right|_{n=n_0}.
\end{split}
\label{ILRt}
\end{eqnarray}
Next, we show that the contribution of the defect to the conserved quantities cancels out the right hand side of (\ref{ILRt}).
From (\ref{GT}), we find
\begin{eqnarray}
\Omega^2=\left(\mathcal{D}^{21}+\mathcal{D}^{22}\Omega^1\right)\left(\mathcal{D}^{11}+\mathcal{D}^{12}\Omega^1\right)^{-1}.
\label{O21}
\end{eqnarray}
Using (\ref{O21}) and using (\ref{DEb}) at $n=n_0$ to eliminate $\mathcal{T}^{11}_2$ and $\mathcal{T}^{12}_2$,
we can write the right hand side of (\ref{ILRt}) as
\begin{eqnarray}
-\left.\left(\mathcal{D}_t^{11}+\mathcal{D}_t^{12}\Omega^1+\mathcal{D}^{12}\left(\mathcal{T}^{21}_1-2\mathcal{T}^{11}_1\Omega^1-\mathcal{T}^{12}_1\left(\Omega^1\right)^2\right)\right)
\left(\mathcal{D}^{11}+\mathcal{D}^{12}\Omega^1\right)^{-1}\right|_{n=n_0}.
\label{rhs1}
\end{eqnarray}
Equation (\ref{LPTBT}) implies the Ricatti equation
\begin{eqnarray}
\Omega_t^1=\mathcal{T}^{21}_1-2\mathcal{T}^{11}_1\Omega^1-\mathcal{T}^{12}_1\left(\Omega^1\right)^2.
\label{ricati2}
\end{eqnarray}
Using this, the expression (\ref{rhs1}) becomes
\begin{eqnarray}
-\left.\left(\mathcal{D}^{11}+\mathcal{D}^{12}\Omega^1\right)_t
\left(\mathcal{D}^{11}+\mathcal{D}^{12}\Omega^1\right)^{-1}\right|_{n=n_0}.
\label{rhs2}
\end{eqnarray}
From (\ref{ID}), we find that $\partial_t I_{defect}(z)$ cancels out (\ref{rhs2}).
Thus (\ref{It}) holds.
\QEDB

Proposition \ref{pro5} implies that the formula (\ref{GI}) provides a generating function for the infinite number of conserved quantities
for the AL system with a defect at $n_0$.
We are able to derive recursively the explicit forms of these conserved quantities by expanding $\Omega^j(n,t,z)$, $j=1,2$, in terms of negative powers of $z$.
Indeed, from (\ref{LPSBT}) we find that $\Omega^j$ satisfies the following difference equation, for $j=1,2$,
\begin{eqnarray}
q^j(n,t)\Omega^j(n,t,z)\Omega^j(n+1,t,z)+z\Omega^j(n+1,t,z)-z^{-1}\Omega^j(n,t,z)-p^j(n,t)=0.
\label{Od}
\end{eqnarray}
We expand $\Omega^j$ in terms of negative powers of $z$ as
\begin{eqnarray}
\Omega^j=\sum_{k=1}^\infty \Omega^j_k(n,t)z^{-k},  ~j=1,2.
\label{Oz}
\end{eqnarray}
By inserting (\ref{Oz}) into (\ref{Od}) and by equating the coefficients of powers of $z$,
we arrive at
\begin{subequations}
\begin{eqnarray}
\Omega_{2k}^j(n,t)=0, \quad k\geq 1,
\label{wja}
\\
\Omega_{1}^j(n,t)=p^j(n-1,t) , ~~\Omega_{3}^j(n,t)=p^j(n-2,t)-p^j(n-2,t)p^j(n-1,t),
\label{wjb}
\\
\Omega_{2k+1}^j(n,t)=\Omega_{2k-1}^j(n-1,t)-\sum_{l+m=2k}\Omega_l^j(n-1,t)\Omega_m^j(n,t),\quad k\geq 1.
\label{wjc}
\end{eqnarray}
\label{wj}
\end{subequations}
Substituting (\ref{Oz}) and (\ref{wj}) into (\ref{GI}) and (\ref{I}),
we finally obtain the infinite set of conserved quantities for the defect system order by order.

\noindent
{\bf Remark 4.} We note that different aspects regarding integrable boundary
conditions and B\"{a}cklund transformations have been studied in \cite{IB,IB1,IB2}.

\end{appendices}

\vspace{1cm}
\small{

}
\end{document}